\documentclass[%
 reprint,
superscriptaddress,
showpacs,preprintnumbers,
 amsmath,amssymb,
 aps,
 prx,
floatfix,longbibliography
]{revtex4-1}

\usepackage{microtype, lmodern}

\usepackage{graphicx}
\usepackage{dcolumn}
\usepackage{bm}
\usepackage[smaller]{acronym}
\usepackage{siunitx}
\usepackage[english]{babel}
\usepackage{mathtools}

\usepackage{csquotes}
\usepackage{hyperref}
\usepackage{xcolor}
\usepackage{xspace}

\newcommand{\DIFdelFL}[1]{}

\newcommand{\pfrac}[3][\empty]{\frac{\partial^{#1}#2}{\partial{#3}^{#1}}}
\renewcommand{\dfrac}[3][\empty]{\frac{\mathrm d^{#1}#2}{\mathrm d{#3}^{#1}}}
\renewcommand\vec[1]{\bm{\mathrm{#1}}}
\renewcommand\top{\textsf{T}}

\begin{document}

\title{Unveiling domain wall dynamics of ferrimagnets in thermal magnon currents: competition of angular momentum transfer and entropic torque}

\author{Andreas Donges} 
\affiliation{Fachbereich Physik, Universit\"at Konstanz, Universit\"atsstra\ss e 10, DE-78457 Konstanz, Germany}%

\author{Niklas Grimm} 
\affiliation{Fachbereich Physik, Universit\"at Konstanz, Universit\"atsstra\ss e 10, DE-78457 Konstanz, Germany}%

\author{Florian Jakobs} 
\affiliation{Fachbereich Physik, Universit\"at Konstanz, Universit\"atsstra\ss e 10, DE-78457 Konstanz, Germany}%
\affiliation{Dahlem Center for Complex Quantum Systems and Fachbereich Physik, Freie Universit\"at Berlin, Arnimallee 14, DE-14195 Berlin, Germany}

\author{Severin Selzer} 
\affiliation{Fachbereich Physik, Universit\"at Konstanz, Universit\"atsstra\ss e 10, DE-78457 Konstanz, Germany}%

\author{Ulrike Ritzmann} 
\affiliation{Fachbereich Physik, Universit\"at Konstanz, Universit\"atsstra\ss e 10, DE-78457 Konstanz, Germany}%
\affiliation{Dahlem Center for Complex Quantum Systems and Fachbereich Physik, Freie Universit\"at Berlin, Arnimallee 14, DE-14195 Berlin, Germany}

\author{Unai Atxitia} 
\affiliation{Fachbereich Physik, Universit\"at Konstanz, Universit\"atsstra\ss e 10, DE-78457 Konstanz, Germany}%
\affiliation{Dahlem Center for Complex Quantum Systems and  Fachbereich Physik, Freie Universit\"at Berlin, Arnimallee 14, DE-14195 Berlin, Germany}

\author{Ulrich Nowak} 
\affiliation{Fachbereich Physik, Universit\"at Konstanz, Universit\"atsstra\ss e 10, DE-78457 Konstanz, Germany}%

\date{\today}

\newacro{LSWT}{linear spin-wave theory}
\newacro{LLG}{Landau-Lifshitz-Gilbert}
\newacro{LLB}{Landau-Lifshitz-Bloch}
\newacro{RE}{rare earth}
\newacro{TM}{transition metal}
\newacro{FI}{ferrimagnet}
\newacro{FM}{ferromagnet}
\newacro{AFM}{antiferromagnet}
\newacro{DW}{domain wall}
\newacro{GPU}{graphics processing units}
\newacro{SSE}{spin Seebeck effect}
\newacro{STT}{spin transfer torque}
\newacro{CUDA}{compute unified device architecture}
\newacro{API}{application programming interface}
\newacro{BZ}{Brillouin zone}
\newacro{YIG}{yttrium iron garnet}
\newacro{GdIG}{gadolinium iron garnet}
\newacro{XMCD}{x-ray magnetic circular dichroism}
\newacro{MOKE}{magneto-optical Kerr effect}
\newacro{HD-AOS}{helicity-dependent all-optical switching}
\newacro{TIMS}{thermally-induced magnetization switching}
\newacro{HAMR}{heat-assisted magnetic recording}
\newacro{DMI}{Dzyaloshinskii-Moriya interaction}
\newacro{WFM}{weak ferromagnet}

\begin{abstract}
Control of magnetic domain wall motion holds promise for efficient manipulation and transfer of magnetically stored information. 
Thermal magnon currents, generated by temperature gradients, can be used to move magnetic textures, from domain walls, to magnetic vortices and skyrmions.  
In the last years, theoretical studies have centered in ferro- and antiferromagnetic spin structures, where domain walls always move towards the hotter end of the thermal gradient. 
Here we perform numerical studies using atomistic spin dynamics simulations and complementary analytical calculations to derive an equation of motion for the domain wall velocity.
We demonstrate that in ferrimagnets, domain wall motion under thermal magnon currents shows a much richer dynamics. 
Below the Walker breakdown, we find that the temperature gradient always pulls the domain wall towards the hot end by minimizating its free energy, in agreement with the observations for ferro- and antiferromagnets in the same regime.  
Above Walker breakdown, the ferrimagnetic domain wall can show the opposite, counterintuitive behavior of moving towards the cold end. 
We show that in this case, the motion to the hotter or the colder ends is driven by angular momentum transfer and therefore strongly related to the angular momentum compensation temperature, a unique property of ferrimagnets where the intrinsic angular momentum of the ferrimagnet is zero while the sublattice angular momentum remains finite. 
In particular, we find that below the compensation temperature the wall moves towards the cold end, whereas above it, towards the hot end.
Moreover, we find that for ferrimagnets, there is a torque compensation temperature at which the domain wall dynamics shows similar characteristics to antiferromagnets, that is, quasi-inertia-free motion and the absence of Walker breakdown.
This finding opens the door for fast control of magnetic domains as given by the antiferromagnetic character while conserving the advantage of ferromagnets in terms of measuring and control by conventional means such as magnetic fields.    

\end{abstract}


\maketitle

\section{Introduction}
Fundamental interest in the understanding of the interaction of thermal stimuli and magnetic domains has been propelled by its potential to impact recording and processing technologies for magnetically stored information \cite{lit:Science-Parkin1, lit:Nat-LLK}.
Control of magnetic states by thermally generated stimuli is hence a growing field of research.
Prominent examples include the fields of spin caloritronics \cite{lit:EES-Boona-2014, lit:Nat-Bauer2012}, e.g. \ac{DW} motion by temperature gradients \cite{lit:PRL-Jiang-etal, lit:PRB-Shokr-etal, lit:PRL-HN, lit:PRL-SRHN, lit:PRL-SARHN, lit:PRB-KT-2015, lit:arxiv-CYQL, lit:PRB-MRML}, 
and the field of ultrafast spin dynamics, e.g. thermally-induced magnetic toggle-switching by ultrafast heat load in \acp{FI} \cite{lit:Nat-Radu-etal, lit:Nat-Ostler-etal, lit:PRB-WHCON,lit:PRB-Wil-etal-1, lit:NL-Liu-etal, lit:PRB-Atx-2013, lit:SR-Bar-etal, lit:APL-AOCC, lit:arxiv-Ger-etal, lit:APL-ElGhazaly-etal}. 
Other coherent means of manipulating magnetic textures, such as \ac{HAMR} \cite{lit:Nat-Stipe2010, lit:Nat-Challener2009} and \ac{HD-AOS} for instance \cite{lit:PRL-Stanciu-etal, lit:Nat-Mangin-etal, lit:PRB-Hass-etal, lit:AM-Hass-etal, lit:PRB-Vahaplar-etal, lit:PRL-Vahaplar-etal}---albeit not primarily induced thermally---are facilitated tremendously by additional application of an ultrashort thermal excitation und subsequent demagnetization \cite{lit:Nat-Bergeard-etal, lit:PRB-Ferte-etal, lit:SPIN-Radu-etal, lit:Nat-Gra-etal, lit:PRL-Radu-etal, lit:PRL-Hennecke-etal}.

A key ingredient for the theoretical description of the aforementioned magnetothermal effects, especially thermally-induced \ac{DW} motion, lies in the understanding of transport processes for energy and angular momentum.
While in metals thermal spin currents are also transported by electrons, in insulators magnons, low energy magnetic excitations, are responsible for the transport of  angular momentum via the \ac{SSE} \cite{Uchida2010b}. 
Notably, thermal magnons can be used to move magnetic textures, such as \acp{DW}, vortices, and skyrmions~\cite{lit:PRL-KZ, lit:PRB-Kovalev, lit:Sky-paper}. 
In previous works the \ac{DW} motion of \acp{FM} and \acp{AFM} induced by temperature gradients has been investigated thoroughly \cite{lit:PRL-HN, lit:PRL-SRHN, lit:PRL-SARHN, lit:PRB-KT-2015, lit:arxiv-CYQL}. 
For instance, both, experimental \cite{lit:PRL-Jiang-etal, lit:PRB-Quessab-etal} and theoretical \cite{lit:PRL-HN, lit:PRL-SRHN, lit:PRB-MRML, lit:PRB-KT-2015} studies on \acp{FM}, have shown that a \ac{DW} in a temperature gradient moves towards the hotter end of the sample.
On a microscopic level, the hot sample region acts as a magnon source.
Since ferromagnetic magnons carry spin, angular momentum conservation dictates that a magnon which is transmitted through a \ac{DW} 
exerts an adiabatic \ac{STT} onto the wall. 
As a consequence, the \ac{DW} moves in opposite direction to the magnon propagation direction, i.e. towards the source \cite{lit:JETP-MY, lit:PRL-HN, lit:PRL-YWW}.
Differently to the mechanism based on angular momentum conservation, an alternative explanation based on thermodynamic arguments has been suggested. 
Since the  \ac{DW}-free energy decreases as the temperature increases \cite{lit:PRB-Hinzke-etal}, the so-called  non-adiabatic entropic torque acts on the magnetization pulling the magnetic texture towards the hotter region of the sample, thereby maximizing the entropy and minimizing the free energy \cite{lit:PRL-SRHN, lit:PRB-KT-2015}.
The generality of the latter picture makes it also applicable to \acp{DW} in \acp{AFM}, in which thermal magnons do on average not carry angular momentum \cite{lit:PRL-SARHN,lit:arxiv-CYQL}, but also to more complex systems such as spin-spirals and skyrmions \cite{lit:PRB-Kovalev}.

Domain wall motion by thermal gradients in \acp{AFM} offers complementary properties to the motion in \acp{FM}. 
On the one hand, \ac{AFM} \ac{DW} motion can be faster due to the almost complete lack of inertia and the missing Walker breakdown, which limits the maximum velocity. 
On the other hand, a disadvantage of \ac{AFM} \acp{DW} is the difficulty to manipulate, control and measure by conventional means, such as external magnetic fields.
This kind of conventional magnetization control is only possible in a subclass of \acp{AFM}, so-called \acp{WFM} such as the \ac{RE}-orthoferrites for instance, in which the \ac{DMI} induces a small net-magnetic moment, perpendicular to the N\'eel order parameter \cite{lit:LTP-Ivanov, lit:SPU-Baryakhtar-1985}. 
So-called \enquote{pure} \acp{AFM}, such as NiO for instance, in which there is no net-magnetization in bulk, require more sophisticated means of excitation \cite{lit:APL-Gomonay-2016, lit:PRL-GJS}.
\acp{FI} can be seen as a generalization of both systems, \acp{FM} and \acp{AFM}, since one may selectively tune the relevant magnetic properties by modifying for instance the sample temperature or composition \cite{lit:Nat-Kim-etal, lit:Nat-Caretta-etal}. 
This allows for an enhanced control of the ferromagnetic- or antiferromagnetic-like character of the spin dynamics and enables to potentially exploit the characteristically fast spin-dynamics of an \ac{AFM} \cite{lit:Nat-Caretta-etal, lit:Nat-Kim-etal, lit:Nat-YRP}, while at the same time one can easily manipulate them by using magnetic fields, and measure it by conventional detection methods such as the \ac{MOKE} or \ac{XMCD}.

Naturally, the larger parameter space of the \ac{FI}, which emerges from the (at least) two non-equivalent magnetic sublattices, also implies that its magnetization dynamics becomes more complex to understand, i.e. the properties of thermal magnon currents strongly depend on the underlying microscopic spin structure \cite{lit:Nat-Gep-etal, lit:PRB-RHN-2017}.
Thus, \ac{DW} motion in \ac{FI} driven by temperature gradients has been scarcely investigated so far \cite{lit:PRB-Shokr-etal} and previous works on \ac{DW} motion in \acp{FI} (and synthetic \acp{AFM}) were focused on more controllable stimuli, such as electric currents \cite{lit:Nat-YRP, lit:PRL-Siddiqui-etal} and magnetic fields~\cite{lit:Nat-Kim-etal}.

In this work we study \ac{DW}-dynamics in \acp{FI} driven by thermal magnon currents in constant temperature gradients \cite{lit:PRB-RHN-2014, lit:PRB-RHN-2017}.
We use an atomistic spin model based on the stochastic \ac{LLG} equation, to simulate ferrimagnetic \acp{DW} in a temperature gradient.
Our simulation results will be compared to the previously developed theory for \ac{DW} motion in \acp{FM}  \cite{lit:PRL-ZL, lit:EPL-TNMS, lit:PRB-Schieback-etal}, based on the collective coordinates approach.
Depending on the strength of the thermal gradient and the base temperature, we find similarities in the \ac{DW} dynamics to both, the \ac{FM} and \ac{AFM}.
For instance we can find a Walker breakdown as observed for \acp{FM} \cite{lit:PRL-SRHN, lit:PRL-HN}, but we also find the quasi-inertia-free motion observed in \acp{AFM} \cite{lit:PRL-SARHN}.
However, in addition we find a completely new feature that is unique to the \ac{FI} and has so far neither been reported for the \ac{FM} nor the \ac{AFM}: a motion towards the cold sample region in the case of a \ac{FI} \emph{below} angular momentum compensation and \emph{above} Walker breakdown.
Using a theoretical model based on \ac{LSWT} we show that this peculiar motion is due to \emph{angular} momentum transfer and not \emph{linear} momentum transfer.

\section{Methods}\label{sec:methods}

\subsection{Atomistic spin model}\label{sec:asm}
\begin{figure}[t!]
	\centering
	\includegraphics[width = \columnwidth]{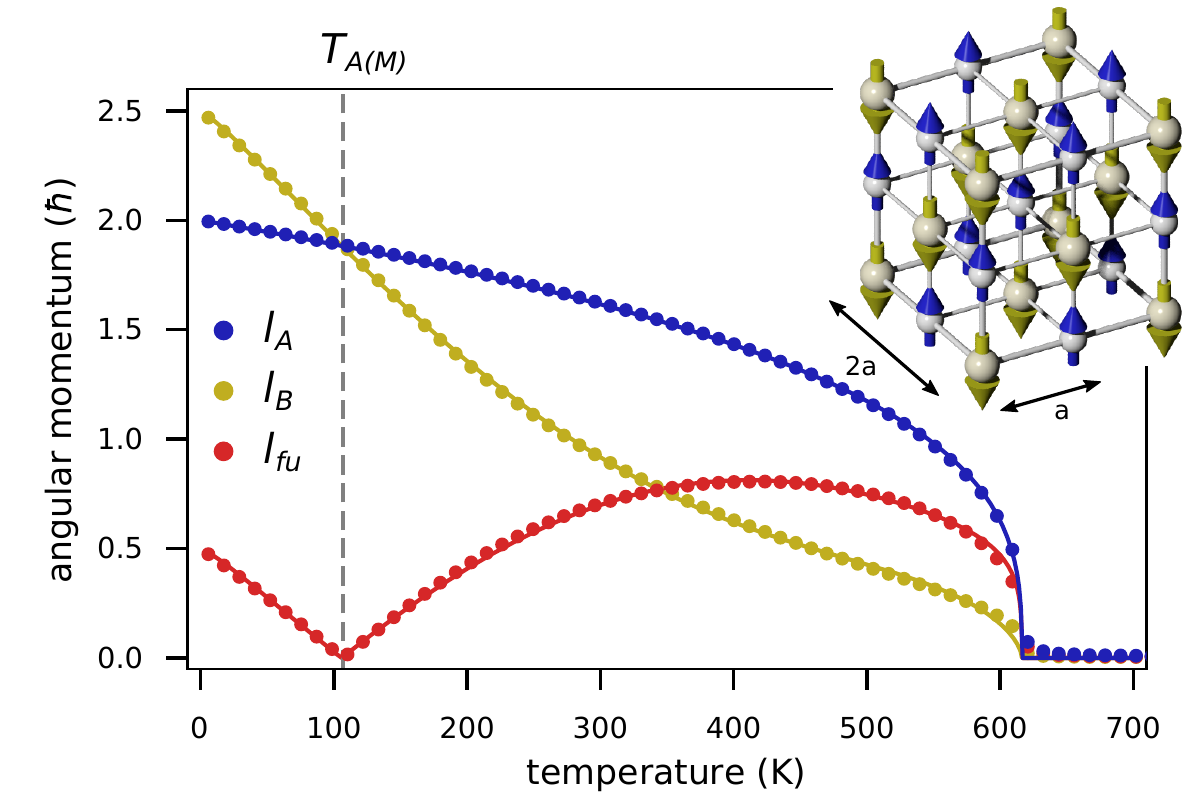}
	\caption
	{
	Sublattice-specific and net thermal average angular momentum as a function of temperature (points). 
	Solid lines are fits to the simulation data. 
	At the angular momentum compensation temperature, $T_{\textrm{A}}$, the net angular momentum vanishes. 
	Note that $T_\text A = T_\text M$, due to our choice of $\gamma_\text A = \gamma_\text B$.
	The sketch shows the G-type magnetic ordering of the underlying atomic spin model of the ferrimagnet.
	}
	\label{fig:curie}
\end{figure}

We model the most simple kind of \ac{FI}, that is a two-sublattice \ac{FI} with a rock salt structure (G-type magnetic ordering) as depicted in Fig.~\ref{fig:curie}.
Our atomistic spin model is based on an extended Heisenberg Hamiltonian 
\begin{gather}
	\mathcal H = -\sum_{i<j}\mathfrak J_{ij} \vec S_i^\top\vec S_j - \sum_i\vec S_i^\top\mathfrak K_i\vec S_i \label{eq:Hamiltonian}
\end{gather}
for normalized magnetic moments $\vec S_i = \vec \mu_i/\mu_i$.
$\mathfrak J_{ij}$ denotes the isotropic Heisenberg exchange coupling and $\mathfrak K_i = d_i^z\vec {\hat z}\vec {\hat z}^\top + d_i^y\vec {\hat y}\vec {\hat y}^\top$ is the biaxial on-site anisotropy with easy $z$-axis and hard $x$-axis ($0\leq d_i^y<d_i^z=\SI{0.5}{\milli\electronvolt}$).
We use the following exchange parameters for the interaction between spins located on the same sublattice A/B,  $J_\text{AA}=\SI{16}{\milli\electronvolt}$, $J_\text{BB}=\SI{0.5}{\milli\electronvolt}$, and between spins on different sublattices, $J_\text{AB}=-\SI{6}{\milli\electronvolt}$. 
These values are characteristic for ferrimagnetic \ac{RE}-\ac{TM} alloys \cite{lit:PRB-Donges-etal}, which are testbed materials in the field of ultrafast spin dynamics \cite{lit:Nat-Radu-etal, lit:Nat-Ostler-etal, lit:PRB-WHCON, lit:PRB-Wil-etal-1, lit:NL-Liu-etal}, and are receiving an increasing attention in the field of spintronics \cite{lit:PRL-Siddiqui-etal}.

The time evolution of the spins is computed with the stochastic \ac{LLG} equation \cite{lit:Now}
\begin{gather}
(1+\alpha^2_\text G)\pfrac{\vec S_i}{t} = -\frac{\gamma_i}{\mu_i}\vec S_i\times\left(\vec H_i + \alpha_\text G\vec S_i\times \vec H_i\right)\label{eq:llg}
\end{gather}
with the effective field $\vec H_i = -\partial\mathcal H/\partial\vec S_i + \vec \zeta_i$, containing both the deterministic field from the spin Hamiltonian $\mathcal H$, Eq.~\eqref{eq:Hamiltonian}, and the stochastic field $\vec \zeta_i$ in the form of Gau{\ss}ian white noise, with 
\begin{gather}
\langle\vec\zeta_i\rangle=0~~\text{and}~~\langle\vec\zeta_i(0)\vec\zeta_j^\top(t)\rangle=\frac{2\alpha_\text G\mu_ik_\text BT_i}{\gamma_i}\delta_{ij}\delta(t).\label{eq:noise}
\end{gather}
For the atomic magnetic moments we use $\mu_\text{A} = 4\mu_\text{Bohr}$ and $\mu_\text{B} = 5\mu_\text{Bohr}$.
Here, for simplicity, we assume that the Gilbert damping and gyromagnetic ratios are the same for both sublattices. 
The gyromagnetic ratios are $\gamma_i=2\mu_\text {Bohr}/\hbar=\SI{1.76e11}{\per\second\per\tesla}$, and the Gilbert damping is set to $\alpha_\text G=0.01$. By numerical integration of Eq. \eqref{eq:llg}, for a range of temperatures we calculate the thermal average of the sublattice-specific  as well as the net angular momentum (Fig.~\ref{fig:curie}). 
The angular momentum compensation temperature, at which the net angular momentum is zero, is found at $T_\text A = \SI{107}{\kelvin}$ (Fig.~\ref{fig:curie}).
Moreover, our numerical calculations allow us to determine the Curie temperature of the system, $T_\text C=\SI{616}{\kelvin}$, in the range of ferrimagnetic \ac{RE}-\ac{TM} alloys \cite{lit:RPP-Buschow, lit:PRB-Zhao-etal}.
We assume a lattice constant of $a=\SI{250}{\pico\meter}$. 
Similar models have been already used in the literature to model the spin dynamics of ferrimagnetic systems, %
the most prominent example being GdFeCo alloys used for ultrafast toggle switching \cite{lit:Nat-Radu-etal, lit:Nat-Ostler-etal, lit:PRB-WHCON, lit:PRB-Wil-etal-1, lit:PRB-Atx-2013, lit:SR-Bar-etal, lit:APL-AOCC} and \ac{HD-AOS}~\cite{lit:PRL-Stanciu-etal, lit:PRB-Vahaplar-etal, lit:PRL-Vahaplar-etal}.
Despite their potential key role on such a switching process, the study of \ac{DW} motion under thermal gradients of such kind of materials \cite{lit:PRB-Shokr-etal} has gained far less attention.

\subsection{Computation of domain wall dynamics}
\begin{figure}[!t]
\centering 
\includegraphics[width = \columnwidth]{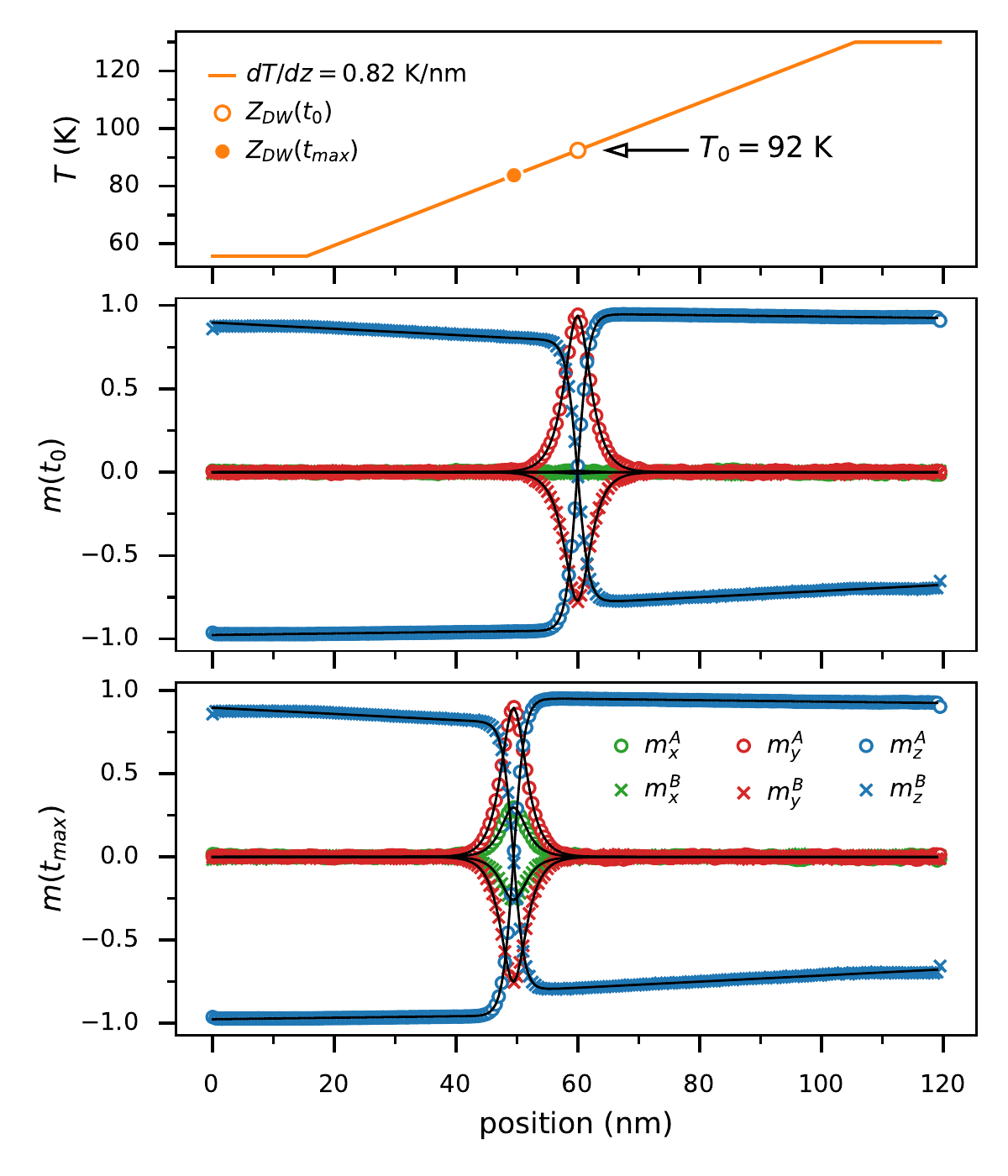}
\caption{Top panel shows the temperature profile used in the simulation setup, together with the initial (solid circle) and final (open circle) \ac{DW} positions; 
$T_0$ indicates the base temperature at the \ac{DW} center at the start of the simulation $t_0$.
Central and bottom panels show the wall profile of the normalized A (circles) and B (crosses) sublattice magnetizations. Black lines correspond to the fits according to Eqs.~\eqref{eq:fit_xy} and \eqref{eq:fit_z}. 
The data shown corresponds to Fig.~\ref{fig:walkers}\,(b).}
\label{fig:profile}
\end{figure}

In our simulations a \ac{DW} is placed in a constant temperature gradient and the magnetization is relaxed to a base temperature of $T_0$.
The base temperature determines the remanent angular-momentum and thus enables us to tune the magnetic properties of the \ac{FI}.
During this relaxation phase ($t<0$) we set $\alpha_\text G=1$ which efficiently suppresses any \ac{DW} dynamics.
At $t=0$ we set $\alpha_\text G$ to $0.01$, which releases the \ac{DW} instantaneously. 
The wall coordinates, i.e. angles $\Phi_\nu$ and positions $Z_\nu$ ($\nu=\text{A,B}$), are tracked by fitting the wall profiles
\begin{align}
m_\nu^\perp(z) &= \frac{m_\nu(z)\exp (i\Phi_\nu)}{\cosh((z-Z_\nu)/\Delta_\nu)},~~\text{and}\label{eq:fit_xy}\\
m_\nu^z(z) &= m_\nu(z)\tanh((z-Z_\nu)/\Delta_\nu)\label{eq:fit_z}
\end{align}
to the simulation data. 
$\Delta_\nu$ is hereby the wall width and $m_\nu(z)$ is the saturation magnetization for which we assume a linear correction in $z$ to improve the fitting accuracy, compared to a spatially constant saturation magnetization.
Absorbing boundaries in the form of enhanced Gilbert damping are applied in the longitudinal direction, whereas the transverse boundaries are periodic in order to have bulk-like properties.
Due to the sizable inter-sublattice coupling $J_\text{AB}$, the deviation of the \ac{DW} coordinates of the two sublattices A and B 
from each other are relatively small, such that for the tracking of the \ac{DW} coordinates it is not necessary to distinguish the \ac{DW} variables $Z,\Phi,\Delta$ for the two sublattices A and B.
A simulation setup of a typical \ac{DW} profile in a temperature gradient is shown in Fig.~\ref{fig:profile}.

We use a comparably large grid cross section of $96\times192$ spins, with a length of \num{480} spins in the direction of the temperature gradient, to reduce thermal fluctuations of the data \cite{foot:grid-size}.
To handle the large computational effort of almost \num{e7} spins in total with simulation times of several hundreds of picoseconds, we use a previously developed, highly efficient \acs{GPU} accelerated atomistic spin dynamics simulation routine based on the \textsc{Nvidia} \acs{CUDA} C-\acs{API}~\cite{lit:cuda-canonical, *lit:cuda101-PG, *lit:cuda101-BPG}.

\section{Theory of domain wall motion}\label{sec:walker-theory}
 \subsection{Adiabatic and non-adiabatic spin transfer torques}

The theory of \ac{DW} motion driven by spin-polarized electric currents in \acp{FM} is well established. 
Initial works have suggested that  thermal magnon currents can be viewed as spin currents, whose amplitude is proportional to the temperature gradient \cite{lit:PRL-SRHN}.  
For \acp{FI}, the question is to what extend a similar picture holds and how theory has to be modified in order to account for the particular properties of \acp{FI}.

In a temperature gradient, the spatial variation of the stochastic noise $\vec \zeta_i(T_i)$, Eq.~\eqref{eq:noise}, in the effective field of the \ac{LLG} equation \eqref{eq:llg} can be interpreted as sources of thermal magnons.
This thermal magnon current acts on a \ac{DW} in a similar way as the \ac{STT} used to describe \ac{DW} motion under spin-polarized electric currents in micromagnetic models \cite{lit:PRL-ZL, lit:EPL-TNMS}.
In those models, the \ac{LLG} equation is augmented by two additional torque terms to take into account the interaction of a spin-polarized electron current on the magnetization. 
The so-called adiabatic torque 
\begin{gather}
\vec T_\text{ad} = {-}(\vec u\cdot\nabla)\vec m \propto -(\nabla T\cdot\nabla)\vec m
\end{gather}
and the non-adiabatic torque 
\begin{gather}
\vec T_\text{nad} = \vec m\times(\beta_\text{eff}\vec u\cdot\nabla)\vec m \propto \vec m\times(\nabla T\cdot\nabla)\vec m.
\end{gather}
The parameter $\beta_\text{eff}$ is the dimensionless non-adiabaticity and by definition specifies the ratio between the two \acp{STT}.

The adiabatic torque can be related to angular momentum conservation \cite{lit:JETP-MY, lit:PRL-YWW, lit:PRL-HN}: when a magnon current (or spin-polarized electron current) passes through the wall, their polarization is continuously rotated by \SI{180}{\degree} thereby changing the magnons' angular momentum.
To obey angular momentum conservation in the combined domain+magnon system, one domain has to grow in size, i.e. the \ac{DW} has to move---the direction depending on the relative polarization of the particle current and magnetization with respect to the \ac{DW}.
The adiabatic torque amplitude in this case is simply given by (cf.~\cite{lit:PRL-YWW})
\begin{gather}
u = Ja^3/l_\text{fu}\label{eq:adiabatic}
\end{gather}
where $J\propto \partial T/\partial z$ (see also Ref.~\cite{lit:PRB-AOTM}  and Appendix \ref{sec:SM:dispersion} and \ref{sec:SM:spin-current}) is the spin current density and $l_\text{fu} = (l_\text A - l_\text B)/2$ the angular momentum per unit volume  ($l_\nu=\mu_\nu m_\nu/\gamma_\nu$), see Fig.~\ref{fig:curie}. The difficulty here is to find an expression for the spin current $J$ for a \ac{FI}.  

On the other hand, \citet{lit:PRL-SRHN} introduced the concept of non-adiabatic entropic torque due to the spatially varying exchange stiffness $\nabla A_\text{eff} = (\partial A_\text{eff}/\partial T) \nabla T $ in a \ac{FM}. 
Here, we adapt their model for the \ac{FM} to the \ac{FI}, and we find the following expression for the non-adiabatic entropic torque strength (see Appendix \ref{sec:SM:Schlickeiser} for the temperature dependence of the exchange stiffness $A_\text{eff}$)
\begin{gather}
\beta_\text{eff} u = -\frac{2a^3}{|l_\text{fu}|}\dfrac{A_\text{eff}}{T}\pfrac Tz.\label{eq:non-adiabatic}
\end{gather}
One of the main result of the present work is the demonstration of the validity of these relations, Eqs. \eqref{eq:adiabatic} and \eqref{eq:non-adiabatic}, for FI, by comprehensive comparison to atomistic spin dynamics simulations of the \ac{DW} motion under a temperature gradient. 
 
 \subsection{Dynamics of the domain wall}

The conceptual idea behind Eqs. \eqref{eq:adiabatic} and \eqref{eq:non-adiabatic} is that the dynamics of a \ac{FI} can be viewed as an effective \ac{FM} with angular momentum given by $l_\text{fu}$---a model which has recently been employed in a similar fashion by \citet{lit:Nat-Kim-etal} for field-driven \ac{FI}-\ac{DW} motion. 
Such a model should be valid for the low wall velocities in thermal gradients; although some deviations might occur close to $T_\text A$ since we do not take into account inertial effects proportional to $\ddot Z$, and $\ddot \Phi$ \cite{lit:JETP-IS, *lit:ZhETF-IS, lit:PRL-GJS}. 

The dynamics of a rigid, ferromagnetic \ac{DW} can then be described by the two collective coordinates that are the wall position $Z_\text{DW}$ and tilting angle $\Phi_\text{DW}$ \cite{lit:EPL-TNMS, lit:PRB-Schieback-etal}. 
We can adapt the corresponding equations of motion and rewrite them for the ferrimagnetic \ac{DW}, leading to 
\begin{align}
\Dot Z_\text{DW} &= \frac{\beta_\text{eff} u}{\alpha_\text{eff}^\perp} -\frac{\Delta_\text{DW}'\dot\Phi_\text{DW}}{\alpha_\text{eff}^\perp}\label{eq:dot-Z}\\
\dot\Phi_\text{DW} &= \frac{\alpha_\text{eff}^\perp}{1+(\alpha_\text{eff}^\perp)^2}\left[\frac{\beta_\text{eff}-\alpha_\text{eff}^\perp}{\Delta_\text{DW}'\alpha_\text{eff}^\perp }u {+} \frac{\mathfrak K_\perp}{|l_\text{fu}|}\sin2\Phi_\text{DW}\right]\label{eq:dot-Phi}
\end{align}
where $\mathfrak K_\perp = \mathfrak K_{yy} - \mathfrak K_{xx}$ is the in-plane anisotropy, $\alpha_\text{eff}^\perp$ is the transverse Gilbert damping parameter, known from magnetic resonance for instance \cite{lit:PRB-SAWHCN, lit:PRB-Kamra-etal}, and $\Delta_\text{DW}'=-\operatorname{sign}(l_\text{fu})\Delta_\text{DW}$ is the signed wall width \cite{foot:wall-width}.
The coupled equations \eqref{eq:dot-Z} and \eqref{eq:dot-Phi} have two kinds of steady state solution.

Below the Walker breakdown \enquote{current} 
\begin{gather}
u_\text W = \frac{\mathfrak K_\perp}{|l_\text{fu}|}\frac{\Delta_\text{DW}\alpha_\text{eff}^\perp}{|\beta_\text{eff} - \alpha_\text{eff}^\perp|}\label{eq:walker-threshold}
\end{gather}
the driving stimulus is insufficient for $\Phi_\text{DW}$ to overcome the potential barrier, scaling with $\mathfrak K_\perp$, hence the wall angle becomes stationary and consequently the motion linear.
In this case, the \ac{DW} velocity reads
\begin{gather}
	V_\text{DW} = \frac{\beta_\text{eff} u }{ \alpha_\text{eff}^\perp} =-\frac{4a^3}{\alpha_\text G|l_\text A + l_\text B|}\dfrac{A_\text{eff}}{T}\pfrac Tz,\label{eq:v-below}
\end{gather} 
where in the second equality we used the relation  \cite{lit:Gur-Mel}
\begin{gather}
\alpha_\text{eff}^\perp \approx \alpha_\text G\frac{|l_\text A + l_\text B|}{|l_\text A - l_\text B|},\label{eq:damping}
\end{gather}
commonly derived under magnetic resonance conditions and related to its linewidth in \acp{FI}. 
We note that Eq. \eqref{eq:v-below} is a generalization of the already known relations for the velocity of the \acp{DW} for \acp{FM} \cite{lit:PRL-SRHN} and \acp{AFM} \cite{lit:PRL-SARHN}. 
Importantly, the difference to those cases is that the effective damping in \acp{FI} has a non-monotonous behavior. 
The relatively simple expression in Eq.~\eqref{eq:damping} holds if the magnetization is not too close to the compensation point, at which an apparent divergence occurs, and where more sophisticated damping models would be necessary \cite{lit:PRB-SAWHCN, lit:PRB-Kamra-etal}.
Thus, very close to this point we expect some deviations in the wall velocity and precession which we derive from the ferromagnetic model here in this section. 
Though qualitatively, this singular behavior accelerates the spin dynamics around $T_\text A$ and is the reason why \acp{FI} with angular momentum compensation points are becoming so relevant for applications and functionalities related to the speed of the spin dynamics \cite{lit:Nat-YRP, lit:Nat-Kim-etal, lit:PRL-Siddiqui-etal}. 
Additionally, while the micromagnetic exchange stiffness and its temperature dependence for \acp{FM} and \acp{AFM} is somehow known \cite{lit:PRB-Atx-etal}, the specifics of $A_{\textrm{eff}}$ in \acp{FI} remains an open problem---especially at elevated temperatures.  

Accordingly, above the Walker breakdown, the wall angle $\Phi_\text{DW}$ precesses continuously as the repelling force is limited by $\mathfrak K_\perp/|l_\text{fu}|$.
The corresponding wall velocity then reads \cite{lit:PRB-Schieback-etal}
\begin{gather}
V_\text{DW} =\frac{\beta_\text{eff} u }{ \alpha_\text{eff}^\perp} {\pm  \frac{|\beta_\text{eff}-\alpha_\text{eff}^\perp|\sqrt{u^2-u_\text W^2}}{\alpha_\text{eff}^\perp (1+\alpha_\text{eff}^{\perp\,2})} }\label{eq:v-above}
\end{gather}
where the positive sign is for the case $(\beta_\text{eff}-\alpha_\text{eff}^\perp)u<0$.

In the limit of a vanishing Walker threshold $u_\text W$ ($\sim \mathfrak K_\perp$) [Eq. \eqref{eq:walker-threshold}] or high driving currents $u \gg u_\text W$, this solution converges to
\begin{gather}
\lim_{u_\text W\to0}V_\text{DW} = \lim_{u\to\infty}V_\text{DW} = \frac{1+\beta_\text{eff}\alpha_\text{eff}^\perp}{1+\alpha_\text{eff}^{\perp\,2}}u\label{eq:v-uniaxial}. 
\end{gather}
Interestingly, the velocity in this limit and in the small damping regime $\alpha_\text{eff}^\perp\ll1$ can be approximated by the simple relation, $V_\text{DW}\approx u = J a^3/l_{\textrm{fu}}$.
We note that the small damping condition, $\alpha_\text G\ll|l_\text A-l_\text B|/|l_\text A+l_\text B|$ holds in a wide range of temperature, when the system temperature is not too close to the compensation temperature.

To evaluate the equations presented in this section, we want to refer to Appendix \ref{sec:SM:dispersion}-\ref{sec:SM:Schlickeiser}.
First of all, in Appendix \ref{sec:SM:dispersion} we derive the dispersion relation of the \ac{FI}, from the \ac{LLG}-equation. 
In Appendix \ref{sec:SM:spin-current} this dispersion relation is used to calculate the thermal spin current density $J$, from which we can then derive the adiabatic \ac{STT} parameter $u$.
Finally, in Appendix \ref{sec:SM:Schlickeiser} we compute the non-adiabatic \ac{STT} for this system using the effective exchange stiffness of the \ac{FI}.
Altogether this allows us to compute the effective \ac{DW} velocity and precession in a self-consistent way from the \ac{LLG}-equation, \eqref{eq:llg}, and the spin Hamiltonian, \eqref{eq:Hamiltonian}, without the need of additional parameters.

\begin{figure}[t!]
	\centering
	\includegraphics[width = \columnwidth]{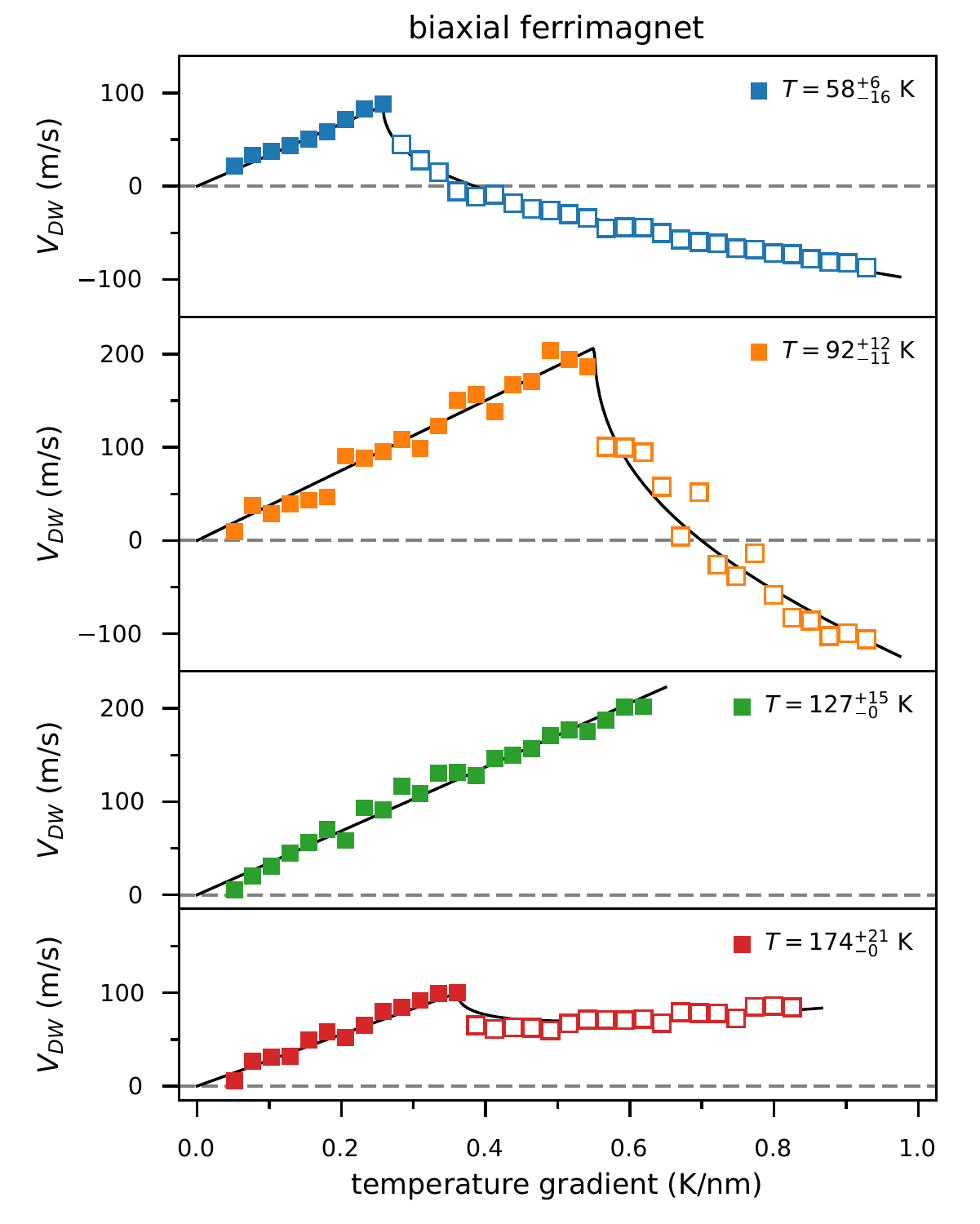}
	\caption
	{
		\ac{DW} velocity as a function of temperature gradient for various temperatures.
		Filled (open) symbols denote walls below (above) Walker breakdown. 
		Error margins indicate the maximum thermal drift of the wall towards the hot (plus) and cold (minus) sample regions throughout the simulation time, i.e. the maximum deviation from the starting temperature.
		The intermediate anisotropy is $d^y_\text A=\SI{12.5}{\micro\electronvolt}$.
		Black lines are fits to Eqs.~\eqref{eq:v-below} and \eqref{eq:v-above}.
	}
	\label{fig:walkers}
\end{figure}

\section{Atomistic Spin Dynamics Simulations}
\subsection{Overview of the domain wall dynamics}

Figure \ref{fig:walkers} shows the wall velocity $V_\text{DW}$ we obtained from our atomistic spin dynamics simulations, as a function of the temperature gradient $\partial T/\partial z$, i.e. the amplitude $u$ of the driving \ac{STT} (the calculation of the steady-state wall velocity is described in detail in Appendix~\ref{sec:SM:steady-state}).  
We study a range of temperature gradients such that we can investigate the dynamics below and above the Walker breakdown.  
Additionally, we consider four different base temperatures $T_0$, ranging from below to above the compensation temperature $T_{\textrm{A}}=\SI{107}{\kelvin}$ that will allow us to assess the role of the net angular momentum  in the dynamics of the DW.

Below the Walker breakdown (filled symbols) we find that the wall velocity $V_\text{DW}$  scales linearly with the 
 thermal torque amplitude $u$ as it is expected according to Eq.~\eqref{eq:v-below}.
The positive sign of the velocity hereby indicates a motion towards the hotter end as previously predicted for \acp{FM} and \acp{AFM} \cite{lit:PRL-SRHN, lit:PRB-KT-2015, lit:PRL-SARHN, lit:PRL-HN}.
Above the Walker breakdown the situation is vastly different:
here we find that below the compensation point the wall moves towards the cold region ($V_\text{DW}<0$) in the limit $u\gg u_\text W$, and above the compensation point to the hot one ($V_\text{DW}>0$) for any value of $u$.
Nevertheless, the theoretical wall velocity predicted by Eqs.~\eqref{eq:v-below} and \eqref{eq:v-above} nicely traces our simulations results for all four temperatures displayed in Fig.~\ref{fig:walkers}.
Thus, different propagation directions of the \ac{DW} motion are found for temperatures below and above the angular momentum compensation temperature $T_\text A$.
This implies that  around $T_\text A$ (i) the adiabatic torque parameter $u(T)$ changes sign  due to the sign change of $J/l_\text{fu}$ in Eq.~\eqref{eq:adiabatic} and (ii) the non-adiabaticity $\beta_\text{eff}(T)$ changes sign likewise  since the product $\beta_\text{eff}u$, Eq.~\eqref{eq:non-adiabatic}, is strictly positive (for $\partial T/\partial z>0$). 

Another intriguing observation is the apparent increase in the Walker threshold $u_\text W$ for the upper three panels (a)-(c), where in (c) the threshold was actually too high to be determined by our simulations.
This is highly counterintuitive since the critical current in a biaxial magnet is determined by the in-plane anisotropy $\mathfrak K_\perp$ [see Eq.~\eqref{eq:dot-Phi}] which decreases quickly with temperature $\langle\mathfrak K_\nu\rangle (T)\sim \mathfrak K_\nu(0)m^3_\nu(T)$ \cite{lit:JPCS-CC}. 
Thus, the Walker threshold $u_\text W(T)$ is expected to decrease monotonically with $T$ likewise which we observe only at even higher temperatures shown in panel (d).
However, in the \ac{FI}, the expression  $u_\text W\propto\mathfrak K_\perp / |l_\text{fu}|$ in Eq.~\eqref{eq:walker-threshold} increases, since the net-angular momentum $l_\text{fu}(T)$ decreases faster than the individual sublattice order parameters $m_\nu(T)$ and hence faster than $\langle\mathfrak K_\perp\rangle(T)$. 
These insights about the Walker threshold could have impact in the design of ferrimagnetic devices with improved functionalities, as the temperature dependence of the individual order parameters can be readily tuned by material engineering techniques, e.g. by modification of sample composition.

\begin{figure*}[!t]
	\centering
	\includegraphics[width = \textwidth]{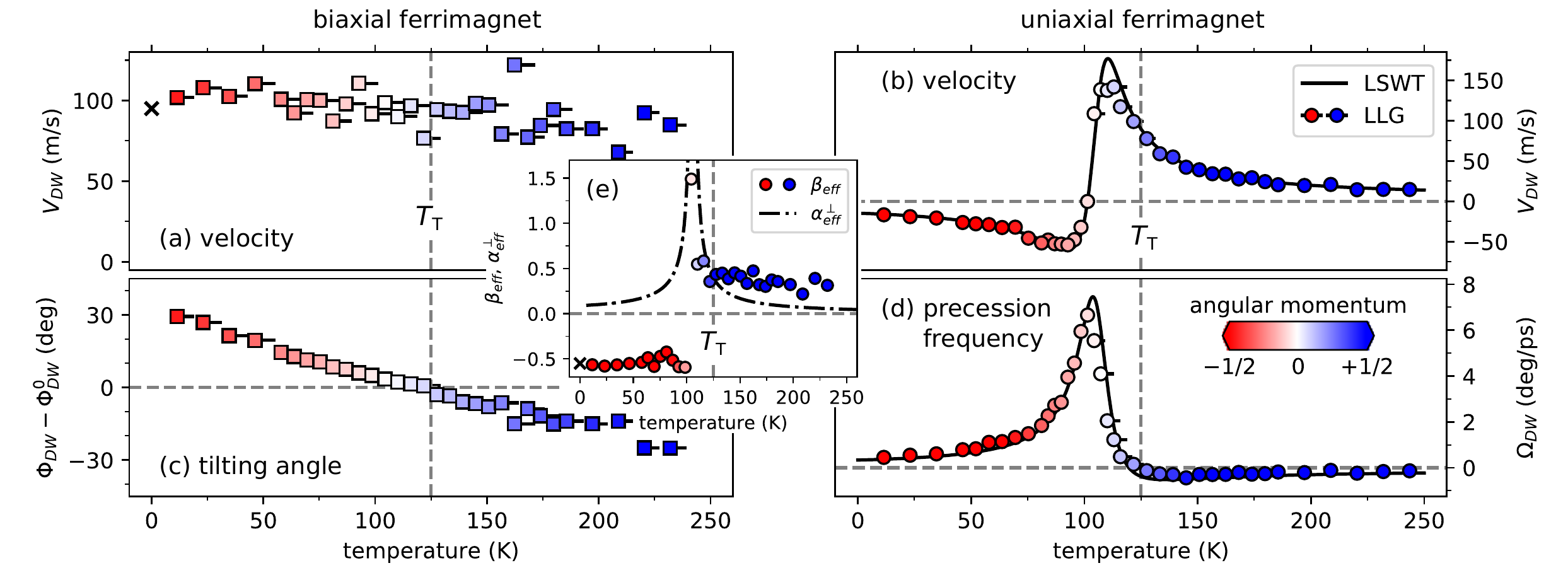}
	\caption
	{
		Top panels show the \ac{DW} velocity $V_\text{DW}$ for a biaxial (a) and a uniaxial (b) \ac{FI} as a function of temperature for a constant thermal gradient of $\partial T/\partial z = \SI{260}{\kelvin\per\micro\meter}$.
		Bottom panels show the corresponding tilting angle of the wall (c) for the biaxial \ac{FI} below Walker breakdown and the \ac{DW} precession frequency for the uniaxial magnet (d), respectively. 
		The inset (e) shows the non-adiabaticity parameter $\beta_\text{eff}$ (markers) calculated from Eq.~\eqref{eq:beta} in comparison to the effective damping coefficient $\alpha_\text{eff}^\perp$ (dash-dotted line), corresponding to the magnetic resonance linewidth, Eq.~\eqref{eq:damping}.
		The black crosses in (a) and (e) correspond to the entropic torque \eqref{eq:non-adiabatic}.
		Solid black lines in (b) and (d) correspond to Eqs.~\eqref{eq:dot-Phi} and \eqref{eq:v-uniaxial}, using the spin current Eq.~\eqref{eq:integral-grad} as input and a constant non-adiabaticity parameter $|\beta_\text{eff}|=\num{0.55}$ (see text).
		Dashed vertical lines mark the so-called torque compensation temperature $T_\text T$ (see text).
		The marker color indicates the net-angular momentum $l_\text{fu}$; horizontal error bars indicate the wall drift.
	}
	\label{fig:temperature-sweep}
\end{figure*}

\subsection{Diversity of temperature dependence of the domain wall dynamics}

 The \ac{DW} dynamics below the Walker breakdown ($u<u_\text W$) behave as one would expect from previous works in \ac{FM} but with effective parameters accounting for the fact that the \ac{FI} is composed of two antiferromagnetically coupled sublattices. 
Thus, it is worth to further investigate the range of validity of this idea. 
As shown above, the Walker threshold can be controlled via the perpendicular anisotropy parameter, $d^y_\text A$. 
Therefore, in order to investigate the regime below Walker breakdown ($u<u_\text W$) we have to consider systems with biaxial anisotropy. 
On the other hand, one of the main results of this work is the demonstration that a \ac{DW} motion towards the cold end of the sample is possible above the Walker breakdown ($u\gg u_\text W$).
Since for a uniaxial \ac{FI} ($d^y_\nu=0$) the wall motion is always above the Walker breakdown ($u_\text W=0$), we can investigate the validity of Eq.~\eqref{eq:v-uniaxial} for the wall velocity without mixing effects coming from the presence of perpendicular anisotropy. 
To study the \ac{DW} dynamics below and above the Walker breakdown more thoroughly, we computed the temperature dependence of the steady state \ac{DW} velocity for the biaxial ($d^y_\text A=\SI{25}{\micro\electronvolt}$) and uniaxial ($d^y_\text A=0$) \ac{FI}, shown in Fig.~\ref{fig:temperature-sweep} (a,b), for a fixed temperature gradient of $\partial T/\partial z= \SI{260}{\kelvin\per\micro\meter}$.

\subsubsection{Domain wall velocity in biaxial systems}

As we could already expect from the data in Fig.~\ref{fig:walkers}, the wall velocity below the Walker breakdown is only weakly sensitive to the base temperature, as can be seen in the panel (a) of Fig.~\ref{fig:temperature-sweep}.
The wall velocity decreases only slightly at higher temperatures. 
However, this effect is already known from previous studies and can be related to the changing equilibrium magnetic properties \cite{lit:arxiv-CYQL, lit:PRB-Atx-etal}.
We can estimate the expected \ac{DW} velocity by evaluating Eq.~\eqref{eq:v-below} (see also Appendix \ref{sec:SM:Schlickeiser}).
For the simulation parameters described in Sec.~\ref{sec:asm}, this yields an \ac{STT} of $\beta_\text{eff}u=\SI{8.6}{\meter\per\second}$ and an expected \ac{DW} velocity of about \SI{95}{\meter\per\second} (black cross) and is in good agreement with the simulation data (colored squares), especially when we consider the rough approximations we applied.
Note that although the entropic torque \eqref{eq:non-adiabatic} alone is proportional to $1/|l_\text{fu}|$ and hence expected to diverge at the angular momentum compensation point (white marker color) where $l_\text{fu}\to0$, the \ac{DW} velocity Eq.~\eqref{eq:v-below} remains finite since the damping coefficient diverges in the same fashion: $\alpha_\text{eff}^\perp\propto 1/|l_\text{fu}|$.

\subsubsection{Domain wall velocity in uniaxial systems}
More fascinating dynamics can be found for the uniaxial \ac{FI}, where predominantly the adiabatic \ac{STT} drives the \ac{DW}, see Fig.~\ref{fig:temperature-sweep} (b).
Here we can make two clear observations: (i) the direction of motion of the wall changes sign very close to the compensation temperature (white marker color).
Below the compensation point the wall moves to the colder sample regions, i.e. co-propagates with the magnon current, whereas above the compensation point the regular motion towards the thermal source is obtained. 
(ii) The absolute value of the wall velocity drastically increases towards the compensation point, which is supported by Eq.~\eqref{eq:v-above} for a magnon current density $J$ which depends only weakly on temperature.
This assumption, in particular that $J$ does not change sign at $T_\text A$, is hereby motivated by its derivation from the dispersion relation (see Appendix~\ref{sec:SM:dispersion}), which does not depend on temperature in a qualitative way, as long as $T\ll T_\text C$~\cite{lit:PRL-BB}.

However, to fully understand the \ac{DW} dynamics for the freely precessing wall, far above the Walker breakdown, we first need a better understanding of the origin of the spin current density $J$.
Until now, it is not even clear what the sign of the spin current density $J$ is in the \ac{FI} \cite{lit:Nat-Gep-etal}.
For a single sublattice \ac{FM} the magnetic moment of the magnon is given by the reduction of the $S_z$ spin component with respect to the saturation value ($m_z=\pm1$) and is hence antiparallel to the ground state magnetization \cite{lit:JETP-MY, lit:PRL-HN}.
In the \ac{FI} the net-momentum of the magnon will be determined by the ratio of the two components $\mu_\text AS_\text A^z/\gamma_\text A$ and $\mu_\text{B}S_\text B^z/\gamma_\text B$ relative to each other.

The classical spin-wave amplitudes for the uniaxial \ac{FI} at $T=\SI0\kelvin$ can be calculated from \ac{LSWT}, following Refs.~\cite{lit:PRB-RHN-2014, lit:JPD-Cramer-etal} (see Appendix \ref{sec:SM:dispersion}).
We find that the low-frequency branch ($\sigma=-1$) in the dispersion relation carries momentum parallel to $m_\text A^z=+1$ and the high-frequency branch ($\sigma=+1$) parallel to $m_\text B^z=-1$.
The question now is which of these two branches dominates the net spin current?
Lower frequency implies higher thermal population---in the classical model that is a population according to a Rayleigh-Jeans distribution $n^0_{\vec k, \sigma} = k_\text BT/\hbar\omega_{\vec k,\sigma}$---and at the same time longer life times $\tau_{\vec k,\sigma}$.
However, the high-frequency branch has a much steeper dispersion relation and hence much higher group velocities $\vec v_{\vec k,\sigma} = \partial\omega_{\vec k,\sigma}/\partial\vec k$ and propagation lengths $\xi_{\vec k,\sigma}=|\vec v_{\vec k,\sigma}|\tau_{\vec k,\sigma}$.

\begin{figure*}[!t]
	\centering
	\includegraphics[width = \textwidth]{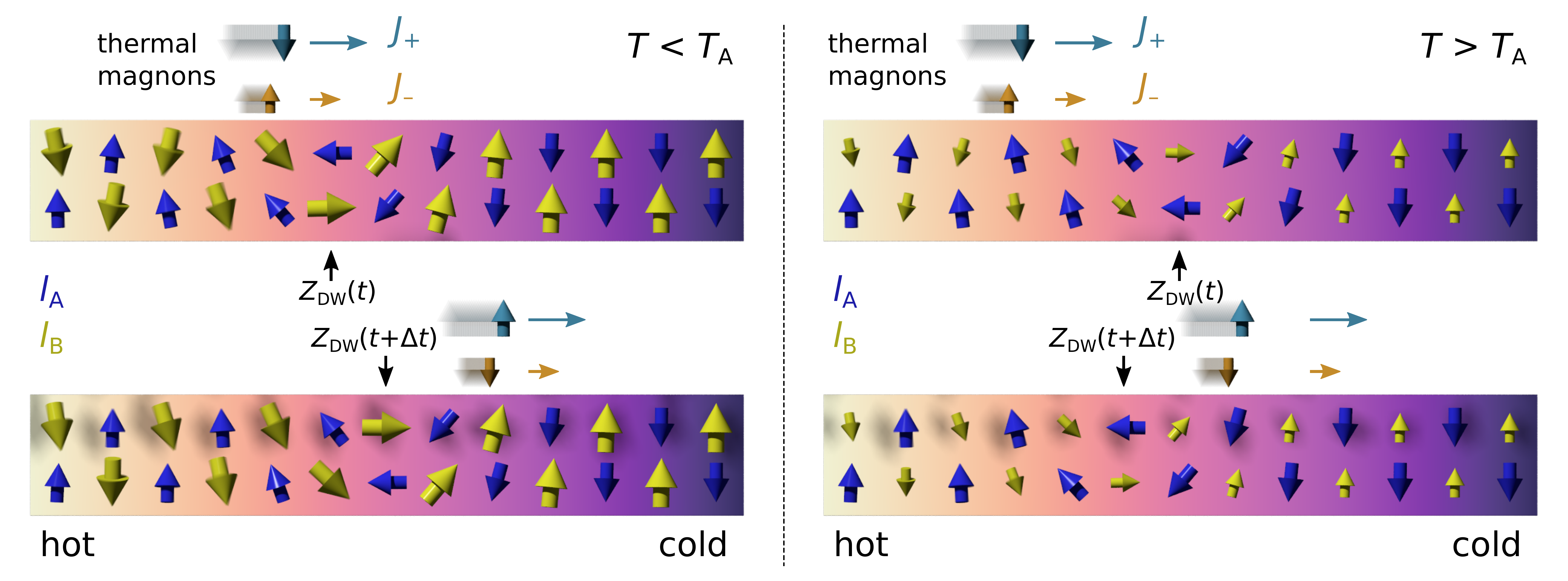}
	\caption
	{
		Schematics of \ac{DW} motion in a thermal gradient due to a magnon current $J_{\sigma}$ for base temperatures $T<T_\text A$ (left) and $T>T_\text A$ (right).
		The magnon current density $J_\sigma$ of the two branches ($\sigma=\pm1$) of the dispersion relation carry opposite angular momentum.
		We find that for both base temperatures $T$, the high frequency branch $\sigma=+1$ (light blue) dominates over the low frequency branch $\sigma=-1$ (orange), see Appendix \ref{sec:SM:dispersion}.
		This leads to a polarization of the net-magnon current density $J=J_++J_-$ parallel to the $l_\text B$ sublattice angular momentum (in the source region), i.e. $|J_+|>|J_-|$.
		Due to the change in the magnon polarization when passing through the wall, to satisfy angular-momentum conservation in the combined domain+magnon system, the domain with the net-angular momentum $l_\text{fu}$ in the \enquote{down} direction has to grow in size.
		Below $T_\text A$ this is the domain on the hot side (left), whereas above $T_\text A$ it is the domain on the cold magnon side (right).
		Hence, for the same spin current density $J$, we obtain different \ac{DW} propagation directions above and below $T_\text A$.
	}
	\label{fig:sketch-wall-motion}
\end{figure*}

To answer this question, we calculate the thermal magnon current density $J$ quantitatively, by solving the following $\vec k$-space integral (see Appendix \ref{sec:SM:dispersion} and \ref{sec:SM:spin-current} for the derivation):
\begin{gather}
 J = \sum_{\mathclap{\sigma=\pm1}}k_\text B\pfrac{T}{z}\sigma\int\displaylimits_{\mathclap{v_{\vec k}^z>0}}\frac{\text d^3k}{(2\pi)^3}\pfrac{\ln\omega_{\vec k,\sigma}}{k_z}\xi_{\vec k,\sigma}\cos\vartheta_{\vec k}.\label{eq:integral-grad}
\end{gather}
The dispersion relation $\omega_{\vec k,\sigma}$ and the magnon propagation lengths $\xi_{\vec k,\sigma}$ can be written in closed form expression, and $\vartheta_{\vec k}$ is simply the angle between the $\vec k$-vector and the $z$-direction.
Thus the numerical solution of Eq.~\eqref{eq:integral-grad} poses only minimal computational effort, and  we find that the high-frequency branch of the dispersion clearly dominates the net-magnon current density $J$.
Thus, we have the peculiar situation in which below the compensation point $T_\text A$, the net-magnon current has a polarization parallel to the ground state angular momentum $l_\text{fu}$---a situation opposite to the case of a  simple \ac{FM}---leading to the opposite direction of \ac{DW} motion, that is, towards the cold sample regions for $T<T_\text A$ (see Fig.~\ref{fig:sketch-wall-motion} left).
We can use the spin current Eq.~\eqref{eq:integral-grad} to compute the adiabatic \ac{STT} quantitatively, yielding $u=\SI{-15.7}{\meter\per\second}$ at $T=0$. 
Moreover, we can now calculate the non-adiabaticity parameter $\beta_\text{eff}$ using our previously determined non-adiabatic torque parameter $\beta_\text{eff}u = +\SI{8.6}{\meter\per\second}$ yielding $\beta_\text{eff} = -\num{0.55}$.
In combination with Eq.~\eqref{eq:v-uniaxial} we can now predict a wall velocity of about $V_\text{DW}(T=0) = \SI{-14.8}{\meter\per\second}$ for the \SI{260}{\kelvin\per\micro\meter} thermal gradient at $T=0$, which as already expected, can be well approximated by $V_\text{DW}\approx u$.
This value matches quite well with the results from our atomistic spin dynamics simulation, although we are unable to simulate $T=0$ exactly.

Now that we have gained some insight on the role of the adiabatic \ac{STT} on the \ac{DW} motion at low temperatures, we can address the domain wall velocity in the uniaxial \ac{FI} at finite temperature. 
In particular, the \ac{DW} velocity in Fig.~\ref{fig:temperature-sweep}\,(b) presents an apparent asymmetry close to the angular momentum compensation point $T_\text A$.
As we discussed above, the magnon current density $J$ above and below $T_\text A$ should not change drastically as it is derived directly from the dispersion relation, cf.~\cite{lit:PRL-BB}.
Thus, we argue that the temperature dependence of $u=Ja^3/l_\text{fu}$ is mostly due to $l_\text{fu}(T)$. 
Consequently, we expect the adiabatic \ac{STT} to act approximately antisymmetrically on the \ac{DW} in the vicinity of $T_\text A$, implying $\min[V_\text{DW}]\approx -\max[V_\text{DW}]$, as indicated in Fig.~\ref{fig:sketch-wall-motion}.
However, the wall velocity from our numerical simulations clearly does not follow this symmetry---we find $\min [V_\text{DW}]= \SI{-54}{\meter\per\second}$ and $\max[V_\text{DW}]=  +\SI{142}{\meter\per\second}$.

We can trace back the origin of such asymmetry by considering Eq.~\eqref{eq:v-uniaxial}
for the domain velocity in the temperature regime close to $T_\text A$. 
We know that both the adiabatic \eqref{eq:adiabatic} and non-adiabatic \acp{STT} \eqref{eq:non-adiabatic} scale via $\sim1/|l_\text{fu}|$, thus, the temperature dependence of $\beta_\text{eff}$ should be mostly due to the softening of the exchange stiffness $A_\text{eff}(T)$, i.e. weak for $T\ll T_\text C$ (cf. Appendix~\ref{sec:SM:Schlickeiser}).
This allows us to assume $|\beta_\text{eff}|\sim\operatorname{const}$ and only include a temperature dependence in the form of the necessary sign change at $T_\text{A}$.
The temperature dependence of $V_\text{DW}(T)$ in Eq.~\eqref{eq:v-uniaxial} is thus due to $\alpha_\text{eff}^\perp(T)$ [Eq.~\eqref{eq:damping}] and $u(T)=Ja^3/l_\text{fu}(T)$ [Eq.~\eqref{eq:adiabatic}]---the former being only relevant close to $T_\text A$.
Using these assumptions we can calculate $V_\text{DW}(T)$ from our theoretical model, shown as the black line in Fig.~\ref{fig:temperature-sweep}\,(b).
This model excellently describes the temperature dependence of the wall velocity over the full temperature range, including its asymmetry.

We conclude that below $T_\text A$, the entropic torque and the angular momentum transfer work against each other, whereas above $T_\text A$ they act in the same direction. 
Naturally, the angular momentum \emph{transfer }becomes less important if angular momentum \emph{conservation }is broken, that is, when $\alpha_\text{eff}^\perp$ becomes large in the vicinity of $T_\text A$.
At the same time, the contribution of the non-adiabatic term $\sim\beta_\text{eff}\alpha_\text{eff}^\perp/(1+\alpha_\text{eff}^{\perp\,2})$ increases.
These findings demonstrate for the first time that a ferrimagnetic \ac{DW} can be pushed away from a \emph{thermal} magnon source by \textit{angular }momentum transfer---an effect which in \acp{FM} and \acp{AFM} can be achieved only by a less efficient \textit{linear }momentum transfer, i.e. magnon reflection~\cite{lit:PRB-MRML, lit:PRB-KTT, lit:PRB-YYS, lit:PRB-Wang-etal}.


\subsection{Emergence of torque compensation temperature}

\begin{figure*}[!t]
	\centering
	\includegraphics[width = \textwidth]{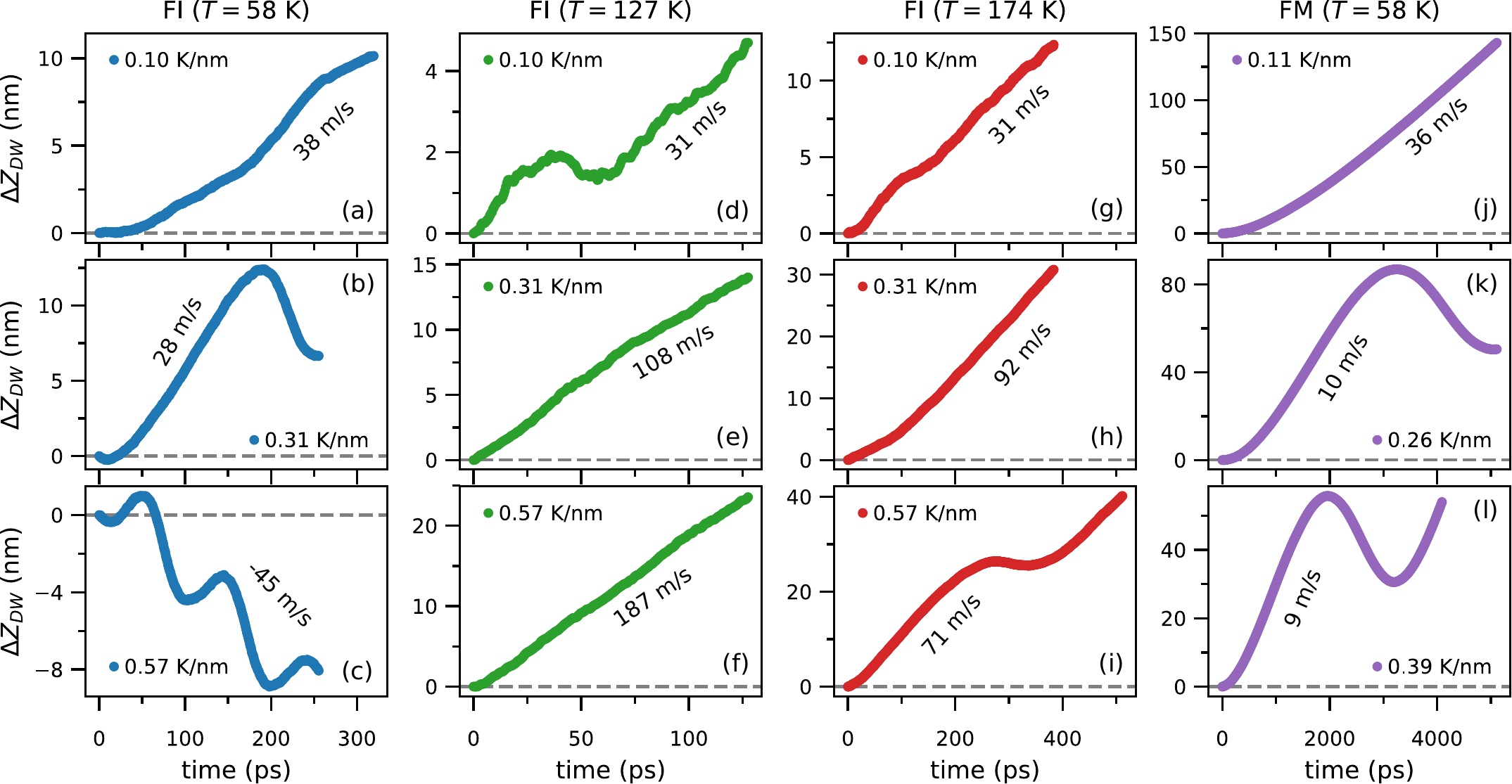}
	\caption
	{
		\ac{DW} displacement $\Delta Z_\text{DW}$ for the biaxial \ac{FI} at different temperatures [(a)-(i); $d^y_\text A=\SI{12.5}{\micro\electronvolt}$] in comparison to the \ac{FM} [(j)-(l); $d^y_\text A=\SI{62.5}{\micro\electronvolt}$].
		Note the much faster acceleration of the ferrimagnetic walls (left and center panels), compared to the ferromagnetic ones (right panels).
		Labels indicate the steady state velocities.
	}
	\label{fig:dynamics}
\end{figure*}

Aside from the \ac{DW} velocity, Fig.~\ref{fig:temperature-sweep} (c) and (d) also show the temperature dependence of the dynamics of both the wall tilting (biaxial) and precession (uniaxial).
For the biaxial \ac{FI} the steady state tilting angle $\Phi_\text{DW}$ gradually decreases with temperature until it reaches zero at a temperature of about \SI{125}{\kelvin} after which it increases again with opposite sense of rotation.
A similar behavior is found for the wall precession $\Omega_\text{DW} = \dot\Phi_\text{DW}$ in the uniaxial \ac{FI}.
The wall precession changes sign at the very same temperature of \SI{125}{\kelvin}, suggesting that this phenomenon is independent of the in-plane anisotropy $\mathfrak K_\perp$.
In the following we define this point of completely suppressed \ac{DW} tilting and precession the  \emph{torque compensation temperature} $T_\text T$.
However, unlike in the biaxial \ac{FI}, in the uniaxial \ac{FI} there is an additional rapid increase of the wall precession frequency $\Omega_\text{DW}$ at the angular momentum compensation point.
Such an increased wall precession close to $T_\text A$ was recently predicted for field-driven \ac{DW} motion in ferrimagnetic GdFeCo by \citet{lit:Nat-Kim-etal}, however, in their case, the sign change of the wall precession coincides with the angular momentum compensation point.
For the case of thermal magnon current driven \ac{DW} motion $T_\text T$ differs from $T_\text A$ implying that field driven \ac{DW} motion is fundamentally different from thermally induced motion.

We can understand the existence of the torque compensation by comparison to the \ac{AFM}.
In the \ac{AFM}, the symmetry of the non-adiabatic \ac{STT} can only lead to propagation $V_\text{DW}$ of the wall along the temperature gradient.
A rotation of the wall angle $\Phi_\text{DW}$ on the other hand does not occur, since both sublattices try to rotate in opposite directions, canting the sublattice magnetizations instead of tilting the \ac{DW} angle \cite{lit:PRL-SARHN}.
In the \ac{FI} this is also the case, but unlike in the \ac{AFM} the torques from the two sublattices will in general not have equal magnitude and are thus not fully compensated.

Another explanation of these results is provided by the coupled equations of motions for the collective coordinates $Z_\text{DW}$ and $\Phi_\text{DW}$, Eqs.~\eqref{eq:dot-Z} and~\eqref{eq:dot-Phi}.
Crucial is hereby the role of the non-adiabaticity parameter $\beta_\text{eff}(T)$, shown in Fig.~\ref{fig:temperature-sweep} (e).
In the previous section we already determined a crude estimate of $\beta_\text{eff}\approx\num{0.55}$ by calculating the non-adiabaticity at low temperature from Eq.~\eqref{eq:non-adiabatic}.
We can use this number again, to solve Eq.~\eqref{eq:dot-Phi} for the steady-state precession frequency, shown in Fig.~\ref{fig:temperature-sweep}\,(d), where we find excellent agreement with the simulation results.

A more rigorous approach to compute the non-adiabaticity $\beta_\text{eff}(T)$, including its temperature dependence, can be computed from the ratio $R=V_\text{DW}^\text{biaxial}/V_\text{DW}^\text{uniaxial}$ of the wall velocity $V_\text{DW}^\text{biaxial}$ below Walker breakdown, Fig.~\ref{fig:temperature-sweep}~(a), and the one for the freely precessing wall $V_\text{DW}^\text{uniaxial}$ Fig.~\ref{fig:temperature-sweep}~(b).
By dividing Eqs.~\eqref{eq:v-below} and \eqref{eq:v-uniaxial} we can then solve for $\beta_\text{eff}(T)$ using $R(T)$ from the numerical simulations:
\begin{gather}
\beta_\text{eff} = \frac{\alpha_\text{eff}^{\perp} R}{1+\alpha_\text{eff}^{\perp\,2}(1-R)}.\label{eq:beta}
\end{gather}
The magnonic torque $u$ drives the wall precession via $\Omega_\text{DW}\propto (\beta_\text{eff} - \alpha_\text{eff}^\perp)u$. 
Thus, the \ac{DW} precession is expected to cease for $\beta_\text{eff}(T) = \alpha_\text{eff}^\perp(T)$, or, in other words, the critical gradient $u_\text{W}$, Eq.~\eqref{eq:walker-threshold}, diverges for $\beta_\text{eff}(T) \to \alpha_\text{eff}^\perp(T)$. 
The intersection point $\beta_\text{eff}(T) = \alpha_\text{eff}^\perp(T)$, shown in Fig.~\ref{fig:temperature-sweep} (e), is in very good agreement with the torque compensation point shown in panels (c) and (d) which have been determined directly from wall tilting and precession, respectively.
Note that by definition, this intersection point also marks the temperature at which the wall velocities above and below Walker breakdown, depicted in Figs.~\ref{fig:temperature-sweep} (a) and (b), coincide.
For our parameters that is a steady-state velocity at $T_\text T=\SI{125}{\kelvin}$ of about \SI{90}{\meter\per\second} in both panels.

Furthermore, we are now able to generalize one of our findings, namely that the torque compensation point $T_\text T$ is found above the angular momentum compensation temperature $T_\text A$:
if the adiabatic \ac{STT} mediated by the thermal magnon current $u$ acts repulsive on the \ac{DW}, the non-adiabaticity $\beta_\text{eff}$ has to be negative, to ensure that the non-adiabatic \ac{STT} \eqref{eq:non-adiabatic}, i.e. the product $\beta_\text{eff}u$, remains positive (for $\partial T/\partial z>0$) \cite{lit:PRL-SARHN, lit:PRL-SRHN, lit:PRB-KT-2015}.
Thus, the term $(\beta_\text{eff} - \alpha_\text{eff}^\perp)u$ in Eq.~\eqref{eq:dot-Phi} can only be zero for an attractive adiabatic \ac{STT}, since $\alpha_\text{eff}^\perp$ is strictly positive.

\subsection{Domain wall motion in time domain}

For discussing the \ac{DW} motion in the time domain it is helpful to compare our results on the \ac{FI}'s dynamics to the previous works on \acp{FM} and \acp{AFM} \cite{lit:PRL-SRHN, lit:PRL-SARHN, lit:PRL-HN, lit:arxiv-CYQL, lit:EPL-TNMS, lit:PRB-Schieback-etal}.
For the \ac{FM} we can do this even in a quantitative manner by simply switching the sign of the inter-sublattice coupling $J_\text{AB}$ to get a ferromagnetic exchange between the sublattices A and B.
The steady-state wall velocity $V_\text{DW}$ below the Walker breakdown appears to be more or less unaffected by the sign change of $J_\text{AB}$, as can be seen in the top panel of Fig.~\ref{fig:dynamics}.
This is indeed expected from Eq.~\eqref{eq:v-below} where only the sum of the sublattice angular momenta enters and the effective exchange stiffness $A_\text{eff}(T)$ of the system should be equal in the \ac{FM}, \ac{AFM}, and \ac{FI}.

However, the time to reach this steady-state velocity in the \ac{FM} is greatly extended with acceleration times on the order of nanoseconds [Fig.~\ref{fig:dynamics} (j)-(l)], whereas in the \ac{FI} the \ac{DW} can reach its steady state velocity on time scales of several tens of picoseconds [Fig.~\ref{fig:dynamics} (a)-(c) and (g)-(i)]. 
In fact, close to the torque compensation point [Fig.~\ref{fig:dynamics} (d)-(f)] the acceleration is even faster, though, the exact time constant there is difficult to determine due to strong fluctuations of the \ac{DW} motion even for a grid cross section of $96\times192$ spins.
This is especially problematic for very low gradients as for instance in Fig.~\ref{fig:dynamics} (a), (d), and (g).

The steady-state below the Walker breakdown is characterized by a constant tilting angle $\Phi_\text{DW}$, where the torques of the non-adiabatic \ac{STT} are balanced by the anisotropy torques, see Eqs.~\eqref{eq:dot-Z} and~\eqref{eq:dot-Phi}.
During the initial rotation of the \ac{DW} up to this angle the velocity increases to its steady-state value, and hence one can interpret it as an inertial mass of the wall~\cite{lit:ZNA-D}. 
As mentioned before, these torques are partially compensated in the \ac{FI} greatly reducing the tilting angle and therefore also the effective inertia of the \ac{DW}.
For the same reason the Walker breakdown $u_\text{W}$, at which the wall starts to rotate continuously, is shifted to much higher critical gradients.
At $T=\SI{58}{\kelvin}$ we find a threshold gradient in the \ac{FI} of about $k_\text B|\partial T/\partial z|_\text W^\text{FI}/d_\text{A,FI}^y\approx \SI{1.8}{\per\nano\meter}$ compared to $k_\text B|\partial T/\partial z|^\text{FM}_\text W/d_\text{A,FM}^y\approx \SI{0.15}{\per\nano\meter}$ in the \ac{FM}.
At the torque compensation point the \ac{FI} resembles an \ac{AFM}, for which there is no tilting and hence the wall can move quasi-inertia-free, i.e. without a relevant acceleration time \cite{lit:PRL-SARHN}. 
Ultrafast \ac{DW} acceleration in the \ac{FI} is not only found at exactly the torque compensation point $T_\text T$, but also slightly below, due to the diverging wall precession $\Omega_\text{DW}$ close to the angular momentum compensation point $T_\text A$ [see Fig.~\ref{fig:temperature-sweep}\,(d)].
Thus, even though the wall has to tilt by a finite angle, the steady-state angle is reached on ultrashort time scales of only few picoseconds.

It should be noted though that there are other effects in an \ac{AFM} that can be attributed to a mass of the \ac{DW} \cite{lit:JETP-IS, *lit:ZhETF-IS, lit:PRL-GJS}.
However, these effects are much smaller and proportional to the velocity of the wall, which is here restricted by feasible temperature gradients.

\section{Conclusion}

To summarize our results, we calculated the \ac{DW} dynamics of a \ac{FI} in a thermal gradient using both, large scale atomistic spin dynamics simulations based on the stochastic \ac{LLG}-equation and analytical calculations based on \ac{LSWT}.
Our simulation results are in good agreement with our theoretical findings that we derived from \ac{LSWT}.
Whereas in the thoroughly studied ferromagnetic systems the adiabatic and non-adiabatic \ac{STT} lead qualitatively to the same result \cite{lit:PRL-HN, lit:PRL-SRHN, lit:PRB-KT-2015}---a motion to the hotter sample region---a ferrimagnetic \ac{DW} reacts differently to these two kinds of torques.
The non-adiabatic torque leads to a consistent motion towards the hotter end, as it is the case for the \ac{FM} and \ac{AFM} and can be explained by the free energy minimization via an entropic torque \cite{lit:PRL-SRHN, lit:PRL-SARHN}.
On the other hand the adiabatic \ac{STT} can either push or pull the ferrimagnetic \ac{DW} away from or towards the spin-wave source, depending on whether the temperature is below or above the angular momentum compensation point.
In the \ac{FI} the copropagation of the \ac{DW} with the magnon current at low temperature is not due to \emph{linear} momentum transfer resulting from magnon reflection \cite{lit:PRB-MRML, lit:PRB-KTT, lit:PRB-YYS, lit:PRB-Wang-etal}, but due to the \emph{angular} momentum transfer from the transmitted magnons.
Moreover, the \ac{FI} shows another distinct characteristic point, besides the angular momentum and magnetic compensation points, that is a torque compensation point at which we find a reversal of the \ac{DW} rotation.
Consequently, at the torque compensation point the Walker breakdown is strongly suppressed which suggests that high \ac{DW} velocities and ultrafast \ac{DW} acceleration should be achievable at this point.

Finally, we want to mention that first experimental evidence on copropagation of a \ac{DW} with a thermal magnon current, induced by ultrashort laser pulses, has been reported recently by \citet{lit:PRB-Shokr-etal}.
In their work it is reported that \acp{DW} in a ferrimagnetic GdFeCo alloy will move away from the laser spot center, i.e. against the thermal gradient and towards the cold region, corroborating our findings for the \ac{DW} motion above the Walker breakdown.

\begin{acknowledgments}
  The authors would like to thank Philipp Graus and Johannes Boneberg for fruitful discussions.
  This work was financially supported by the  Deutsche Forschungsgemeinschaft through the SFB~767 \enquote{Controlled Nanosystems}.
  At the FU Berlin support by the Deutsche Forschungsgemeinschaft through SFB/TRR 227 \enquote{Ultrafast Spin Dynamics}, Project A08 is gratefully acknowledged.
  Furthermore, U.R. acknowledges funding from the Deutsche Forschungsgemeinschaft via the project RI 2891/2-1.
\end{acknowledgments}
%


\appendix

\section{Dispersion relation of the rocksalt-type ferrimagnet}\label{sec:SM:dispersion}
We start the derivation of the spin-wave dispersion by introducing the complex vector $\mathcal S = [S_\text A^x + i S_\text A^y, S_\text B^x + i S_\text B^y]$.
This ansatz implies an $x$-$y$-symmetry and thus vanishing in-plane anisotropy $\mathfrak K_\perp = \mathfrak K_{yy} - \mathfrak K_{xx}=0$ in order to avoid dealing with squeezed magnon states~\cite{lit:PRB-KAB}.
We assume a groundstate magnetization of $m_\text A^z = +1$ and $m_\text B^z = -1$.
Following Refs.~\cite{lit:PRB-RHN-2014, lit:JPD-Cramer-etal} one can deduce the linearized \ac{LLG}-equation in $\vec k$-space in analogy to the \ac{FM} and \ac{AFM} case:

\begin{gather}
\pfrac{\mathcal S_{\vec k}}{t} = -\Omega_{\vec k}\cdot\mathcal S_{\vec k}\label{eq:llg-k}
\end{gather}
where the frequency matrix on the right hand side is given by 
\begin{gather}
\Omega_{\vec k} = %
\left[\begin{matrix}
(+i-\alpha_\text G)\Omega_\text{AA}^{\vec k} & (+i-\alpha_\text G)\Omega_\text{AB}^{\vec k}\\
(-i-\alpha_\text G)\Omega_\text{BA}^{\vec k} & (-i-\alpha_\text G)\Omega_\text{BB}^{\vec k}
\end{matrix}\right].\label{eq:C-matrix}
\end{gather}

The matrix elements of Eq.~\eqref{eq:C-matrix} are 
\begin{align}
\Omega_\text{AA}^{\vec k} &= \frac{\gamma_\text{A}}{\mu_\text{A}}\left(6J_\text{AB} - 2d_\text{A}^z  - 2 J_\text{AA}C^{(2)}_{\vec k}\right)\label{eq:Omega_aa}\\
\Omega_\text{AB}^{\vec k} &= \frac{\gamma_\text{A}}{\mu_\text{A}}2J_\text{AB}C_{\vec k}^{(1)}\label{eq:Omega_ab}\\
\Omega_\text{BB}^{\vec k} &= \frac{\gamma_\text{B}}{\mu_\text{B}}\left(6J_\text{AB} - 2d_\text{B}^z - 2 J_\text{BB}C_{\vec k}^{(2)}\right)\label{eq:Omega_bb}\\
\Omega_\text{BA}^{\vec k} &= \frac{\gamma_\text{B}}{\mu_\text{B}}2J_\text{AB}C_{\vec k}^{(1)}\label{eq:Omega_ba}.
\end{align}

The structure factors $C^{(n)}_{\vec k}$ are related to the neighbor positions of shell $n$ and can be expressed as 
\begin{align}
C^{(1)}_{\vec k} & = \sum_\nu\cos (k_\nu a_\nu)\label{eq:C1}\\
C^{(2)}_{\vec k} & = \sum_{\substack{\nu,\kappa\\\nu\neq\kappa}}\left[1-\cos (k_\nu a_\nu)\cos (k_\kappa a_\kappa)\right]\label{eq:C2};
\end{align}
$2a_\nu$ are hereby the lattice constants of the face centered orthorhombic unit cell (see Fig.~\ref{fig:curie}), in the following we assume $a_\nu = a$ for simplicity.

The solution of Eq.~\eqref{eq:llg-k} is given by the eigenvalues of Eq.~\eqref{eq:C-matrix} and can be computed in closed form for the $2\times2$-matrix:
\begin{gather}
\label{eq:Lambda_exact}\Lambda^{\vec k}_\pm =  -\frac{\alpha_\text G-i}2\Omega_\text{AA}^{\vec k} - \frac{\alpha_\text G+i}2\Omega_\text{BB}^{\vec k}\\\nonumber  \mp\frac12\left[\vphantom{\left(\Omega_A^{\vec k}\right)^2} 4(1+\alpha_\text G^2)\left(\Omega_\text{AB}^{\vec k}\Omega_\text{BA}^{\vec k} - \Omega_\text{AA}^{\vec k}\Omega_\text{BB}^{\vec k}\right)\right.  \\\nonumber \left. + \left((\alpha_\text G-i)\Omega_\text{AA}^{\vec k} + (\alpha_\text G+i)\Omega_\text{BB}^{\vec k}\right)^2\right]^{1/2}.
\end{gather}

\begin{figure}[!t]
	\centering 
	\includegraphics[width=\columnwidth]{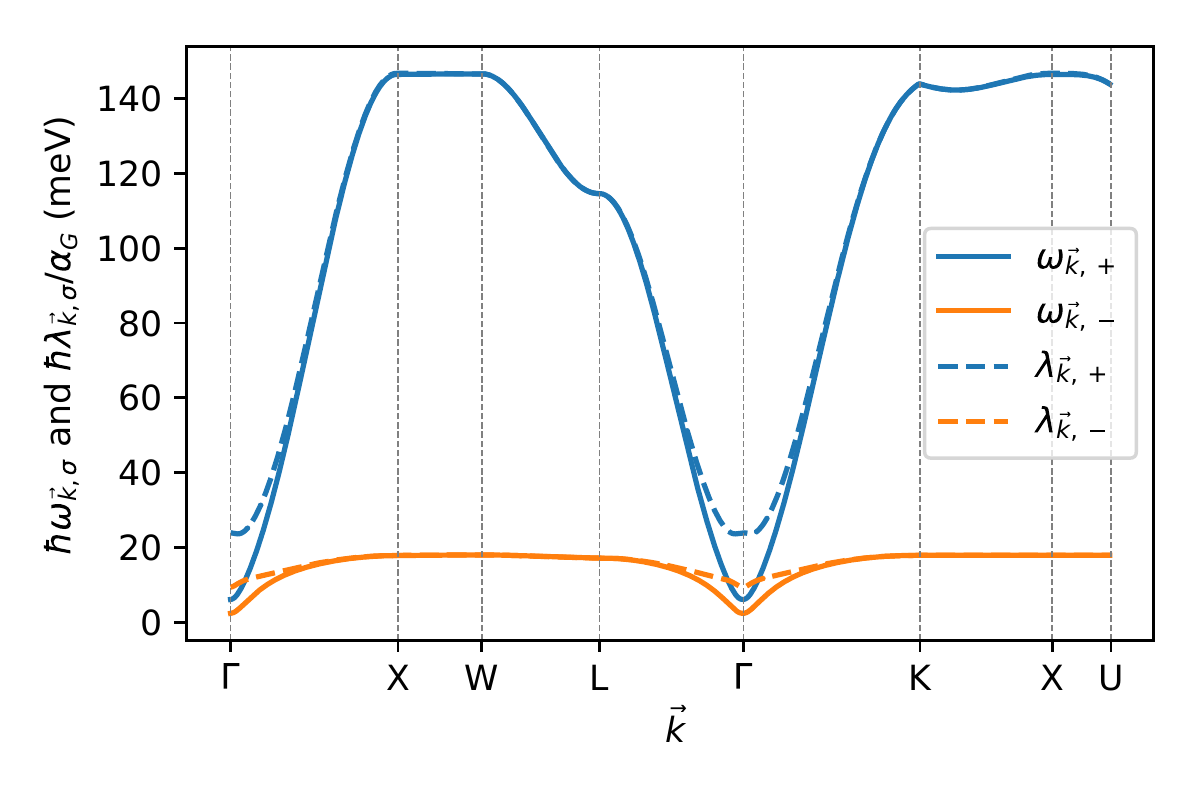}
	\caption
	{
		Dispersion relation of the \ac{FI} along the high symmetry path of the magnetic \ac{BZ} ($\overline{\Gamma\text X}=\pi/a$). 
		Solid lines are the magnon energies and dashed lines their linewidth (normalized to $\alpha_\text G$), respectively.
		Shown are the absolute values of the frequency, neglecting the rotation sense.
	}
	\label{fig:SM:dispersion}
\end{figure}

We can further simplify this expression by assuming $\alpha_\text G\ll1$, leading to 
\begin{align}
 \omega_{\pm}^{\vec k}&= +\frac{\Omega_\text{AA}^{\vec k} - \Omega_\text{BB}^{\vec k} }{2}\mp\frac12\tilde \Omega_{\vec k}\label{eq:omega-k}\\
\frac{\lambda_{\pm}^{\vec k}}{\alpha_\text G} &= {-}\frac{\Omega_\text{AA}^{\vec k} + \Omega_\text{BB}^{\vec k}}2 \pm \frac{(\Omega_\text{AA}^{\vec k})^2 - (\Omega_\text{BB}^{\vec k})^2}{2\tilde \Omega_{\vec k}}\label{eq:lambda-k}
\end{align}
where the frequencies $\omega^{\vec k}_{\pm} = \Im\{\Lambda_{\pm}^{\vec k}\}$ and damping rates $\lambda^{\vec k}_{\pm} = \Re\{\Lambda_{\pm}^{\vec k}\}$ are the imaginary and real parts of the complex eigenvalues $\Lambda_{\pm}^{\vec k}$, respectively. 
The frequency $\tilde\Omega_{\vec k}$ simply reads
\begin{gather}
\tilde\Omega_{\vec k} = \sqrt{(\Omega_\text{AA}^{\vec k} + \Omega_\text{BB}^{\vec k})^2 - 4\Omega_\text{AB}^{\vec k}\Omega_\text{BA}^{\vec k}}.
\end{gather}
Note that for $\vec k=0$ Eqs.~\eqref{eq:omega-k} and \eqref{eq:lambda-k} coincide with the results of \citet[Eq.~(16) and (17)]{lit:PRB-Kamra-etal} for the magnetic resonance mode of a \ac{FI}.

\begin{figure}[!t]
	\centering 
	\includegraphics[width=\columnwidth]{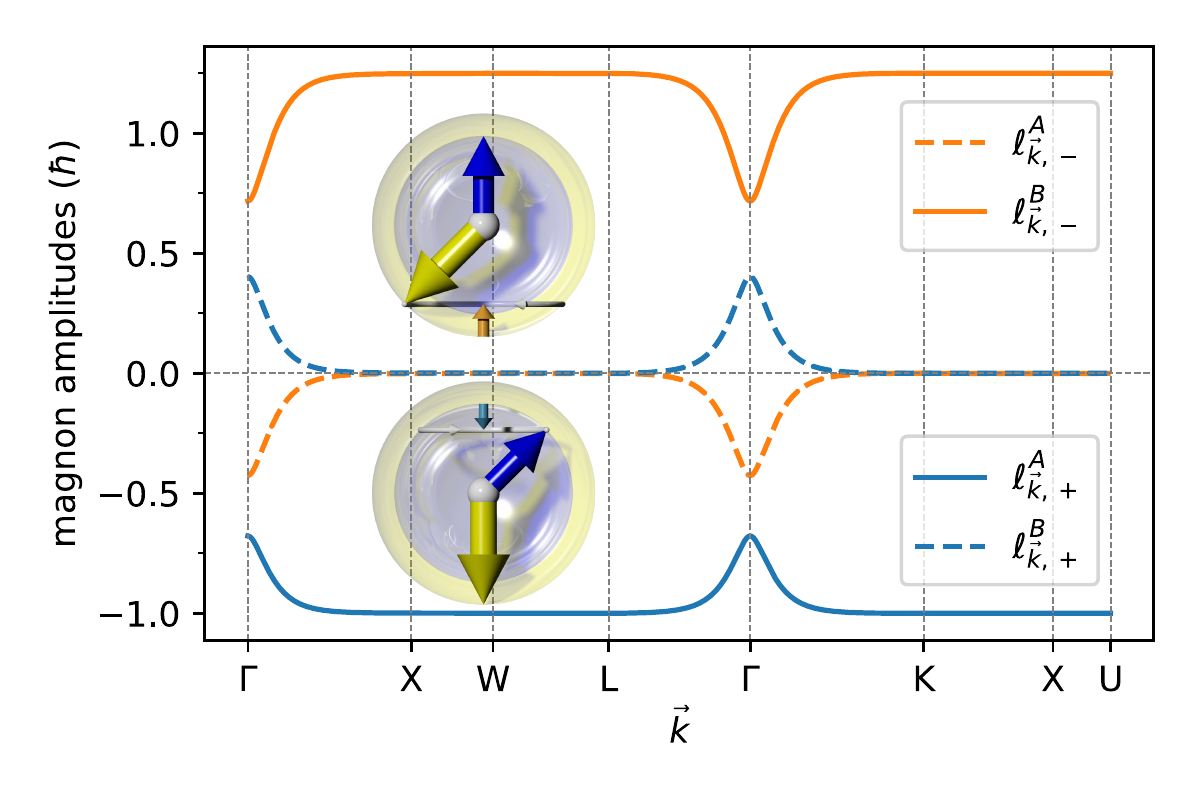}
	\caption
	{
		Sublattice-resolved magnon amplitudes, along the high-symmetry lines in the \ac{BZ}.  
		From the ratio of the amplitudes one can deduce in which direction the net-angular momentum of the magnon points.
		The top/bottom sketch qualitatively show a low-/high-frequency magnon excitation amplitude far away from the $\Gamma$-point.
		At the $\Gamma$-point (not shown as sketch) mixed excitations can occur in which both sublattices are excited.
	}
	\label{fig:SM:amplitudes}
\end{figure}

Next we want to derive the \enquote{amplitude} of the magnon.
Although the absolute value of a magnon is not well defined in our semiclassical picture, it is sufficient to compute the relative amplitudes between the A and B sublattices, as the absolute values of the amplitudes will cancel for a thermal magnon distribution.
These relative amplitudes are related to the (non-normalized) eigenvectors
\begin{gather}
\mathcal S_\pm^{\vec k} = \left[-\frac{\Omega_\text{AA}^{\vec k} + \Omega_\text{BB}^{\vec k}{\mp}\tilde\Omega_{\vec k}}{2\Omega_\text{BA}^{\vec k}}, 1\right],
\end{gather}
to the eigenvalues $\Lambda_\pm^{\vec k}$ of Eq.~\eqref{eq:C-matrix}.
The classical equivalent to the magnon amplitude $\mu_\pm^{\vec k}\ll \mu_\text {A,B}$ follows from simple geometrical considerations as 
\begin{gather}
\mu_\pm^{\vec k} = \frac{{-}S^2}{2||{\mathcal S}_\pm^{\vec k}||^2}\left(\mu_\text A\left({\mathcal S}_{\pm,\text A}^{\vec k}\right)^2 - \mu_\text B\left({\mathcal S}_{\pm,\text B}^{\vec k}\right)^2\right),\label{eq:mu-k}
\end{gather}
or 
\begin{gather}
l_\pm^{\vec k} = \frac{{-}S^2}{2||{\mathcal S}_\pm^{\vec k}||^2}\left(\frac{\mu_\text A}{\gamma_\text A}\left({\mathcal S}_{\pm,\text A}^{\vec k}\right)^2 - \frac{\mu_\text B}{\gamma_\text B}\left({\mathcal S}_{\pm,\text B}^{\vec k}\right)^2\right),\label{eq:l-k}
\end{gather}
where $S$ is a scaling parameter which quantifies the classical spin-wave amplitudes.
The sublattice-resolved magnon amplitudes as defined inside the brackets of Eq.~\eqref{eq:l-k} are shown in Fig.~\ref{fig:SM:amplitudes}. 
One can clearly see that the sign of $\mu_\sigma^{\vec k}$ does not depend on $\vec k$, but only on $\sigma$, since for a given branch $\sigma$, one sublattice is always excited much more strongly.
In fact, apart from the modes close to the $\Gamma$-point, the magnon amplitude can be approximated by an excitation of only one of the two sublattices:
for the low frequency branch (orange), that is the B-sublattice (top), whereas for the high frequency branch (blue) it is the A-sublattice.

\section{Linear spin wave theory for thermally induced domain wall motion}\label{sec:SM:spin-current}
	
\subsection{Temperature step}\label{sec:SM:step}
First we suppose a system of an extended nanostrip with a temperature profile in the form of a step function $T(z) = T_0 + \Delta T~\Theta(-z)$.
The system is then isotropic along the $x$-/$y$-directions and we only expect a net-spin current propagating along $z$-direction.
Since the thermal magnon occupation in the classical limit follows a Rayleigh-Jeans distribution $n_{\vec k,\sigma}^0 = k_\text BT/\hbar\omega_{\vec k,\sigma}$, one can drop the base temperature $T_0$ as long as it is low enough that it does not affect the effective magnetic parameters, i.e. as long as the dispersion relation \eqref{eq:omega-k} and \eqref{eq:lambda-k} is still valid.

Note, that unlike previous works that studied the action of spin-waves on \acp{DW} \cite{lit:PRL-YWW, lit:PRB-KTT}, an effective 1-dimensional model is not sufficient here, since thermally excited magnons with off-axis wave vector $\vec k\neq k\vec{\hat z}$ are relevant and due to the large grid cross-section included in the numerical simulations.
The macroscopic spin current density follows from integrating over all thermally excited modes in the \ac{BZ}.
The two sublattices of the checkerboard \ac{AFM} (Fig.~\ref{fig:curie}) are two fcc-lattices with magnetic lattice constant of $2a$, respectively. 
Thus, the \ac{BZ} is a truncated octahedron with $q_\text{X} = \pi/a$, \cite{lit:JdP-BCFN}.
For a \ac{DW} at a position $z>0$ away from the temperature step, we can restrict the $\vec k$-space integral to the half space $v_{\vec k}^z>0$ to only include forward propagating magnons:
\begin{gather}
  J(z) = \sum_{\sigma=\pm1}\int\displaylimits_{~v_{\vec k}^z>0}\!\frac{\text d^3k}{(2\pi)^3}\, j^z_{\vec k,\sigma}(z)
\end{gather}
Each mode $\vec k,\sigma$ contributes with 
\begin{gather}
j_{\vec k, \sigma} = -\sigma\hbar n_{\vec k,\sigma} \pfrac{\omega_{\vec k, \sigma}}{k_z} \exp\left(\frac{-2z}{\xi_{\vec k,\sigma}\cos\vartheta_{\vec k}}\right)\label{eq:j_k_sigma}
\end{gather}
to the net-current $ J$.
Here $\partial\omega_{\vec k, \sigma}/\partial\vec k = \vec v_{\vec k,\sigma}$ is the group velocity of the mode, $n_{\vec k,\sigma}$ is the magnon occupation number at the source, and the exponential factor accounts for the absorption of the current with propagation length $\xi_{\vec k,\sigma}  = |\vec v_{\vec k,\sigma}|\tau_{\vec k,\sigma}$.
The factor of two in the exponential accounts for the conversion of the spin-wave amplitude to the magnon number, proportional to the squared amplitude.
$\vartheta_{\vec k}$ denotes the angle between $\vec k$ and the $z$-direction and is needed to compute the actual propagated distance $r_{\vec k} = z/\cos\vartheta_{\vec k}$.

\begin{figure}[!t]
	\centering 
	\includegraphics[width = \columnwidth]{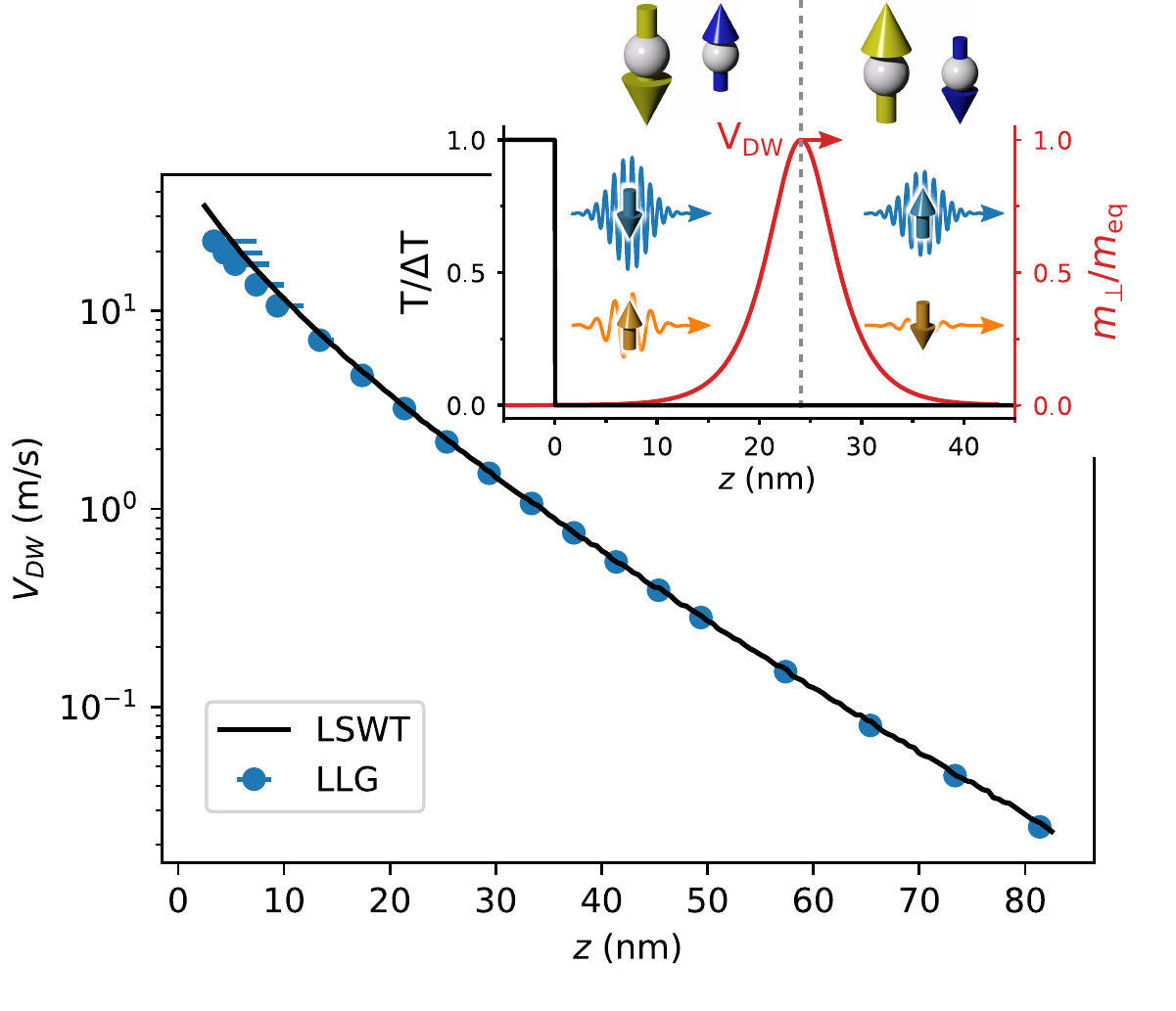}
	\caption
	{
		\ac{DW} velocity for a uniaxial \ac{FI} as a function of distance $z$ from a temperature step of $k_\text BT=\SI{1}{\milli\electronvolt}$; comparison between \acs{LLG}-simulations (points) and \acs{LSWT} (line). 
		The theory line was obtained by integrating Eq.~\eqref{eq:integral-step} numerically via a Monte-Carlo method with about \num{5e5} $\vec k$-points in the $\vec v_{\vec k,\sigma}^z>0$ half of the \ac{BZ}.
		For the \acs{LLG} simulation a grid of $96\times96\times480$ and a simulation time of \SI{192}{\pico\second} was taken ($t\gg z / v^z_{\vec k,\sigma}$ for the majority of the \ac{BZ}).
		Errorbars indicate the initial and final position of the \ac{DW}.
		The \ac{DW} copropagates with the magnon current, i.e. moves to the colder sample part.
		The inset shows the schmematics of the angular momentum transfer due to high (blue) and low (yellow) frequency magnons.
	}
	\label{fig:SM:lswt}
\end{figure}

Ignoring depletion effects at the interface, i.e. at the magnon source, we can assume that the magnon occupation at the source is given by the thermal population $n_{\vec k,\sigma}\approx n_{\vec k,\sigma}^0$ and hence we get
\begin{align}
j_{\vec k, \sigma} &= -\sigma\hbar \frac{k_\text B\Delta T}{\hbar\omega_{\vec k,\sigma}} \pfrac{\omega_{\vec k, \sigma}}{k_z} \exp\left(\frac{-2z}{\xi_{\vec k,\sigma}\cos\vartheta_{\vec k}}\right)\\&=  -\sigma k_\text B\Delta T \pfrac{\ln\omega_{\vec k, \sigma}}{k_z} \exp\left(\frac{-2z}{\xi_{\vec k,\sigma}\cos\vartheta_{\vec k}}\right) 
\end{align}
and we arrive at our preliminary result for the magnon current density due to a temperature step 
\begin{gather}
  J = -\sum_{\mathclap{\sigma=\pm1}}k_\text B\Delta T\sigma\!\int\displaylimits_{\mathclap{v_{\vec k}^z>0}}\!\frac{\text d^3k}{(2\pi)^3}\pfrac{\ln\omega_{\vec k,\sigma}}{k_z}\exp\left(\frac{-2z}{\xi_{\vec k,\sigma}\cos\vartheta_{\vec k}}\right).\label{eq:integral-step}
\end{gather}

We should note that we assumed a fixed magnon amplitude of $\hbar$ in the derivation, which for the quantum mechanical case is a reasonable assumption for the \ac{FI}~\cite{lit:PRB-KAB}, but seems arbitrary for the classical case [see Eq.~\eqref{eq:mu-k}].
This is however not further relevant, since the magnon amplitude eventually cancels when we put in the thermal occupation $n^0_{\vec k,\sigma}\propto1/\hbar\omega_{\vec k,\sigma}$.

In Fig~\ref{fig:SM:lswt} we compare the \ac{DW} velocity calculated according to Eq.~\eqref{eq:integral-step} with our numerical simulations results.
Since the base temperature is set to zero, the system is below the compensation temperature.
The \ac{DW} velocity is plotted as a function of distance $z$ from a \SI{1}{\milli\electronvolt} temperature step. 
We find excellent quantitative agreement between numerical simulations and the \ac{LSWT}.
Furthermore, as for the thermal gradients, we also find the motion of the wall to be away from the magnon source.

\subsection{Temperature gradient}

The solution of the previous section \ref{sec:SM:step} is easily applicable to temperature gradients, by simply summing up over several temperature steps $\mathrm dT(z) = \partial T/\partial z~\mathrm dz$.
Once more we will use the fact that the Rayleigh-Jeans distribution is linear in $T(z)$.
Therefore, in a constant temperature gradient, we have the same amount of magnons flowing from the right to the left (carrying spin $-\sigma\hbar$),  than we have \enquote{magnon-holes} flowing from right to left (carrying spin $+\sigma\hbar$).
This means we can again restrict the $\vec k$-space integral over half of the \ac{BZ} and multiplying the result by two, such that the final result for the spin current is the sum of equal contributions of magnon current and \enquote{magnon-hole} current.
 \begin{gather}
 \mathrm dJ = 2\sum_{\mathclap{\sigma=\pm1}}k_\text B\sigma\int\frac{\text d^3k}{(2\pi)^3}\pfrac{\ln\omega_{\vec k,\sigma}}{k_z}\nonumber\\\times\exp\left(-\frac{2z}{\xi_{\vec k,\sigma}\cos\vartheta_{\vec k}}\right)\,\pfrac{T}{z}\,\mathrm dz.
 \end{gather}
 We obtain the final result Eq.~\eqref{eq:integral-grad} by performing the $z$-integration
 \begin{gather}
 J = \sum_{\mathclap{\sigma=\pm1}}k_\text B\pfrac{T}{z}\sigma\int\frac{\text d^3k}{(2\pi)^3}\pfrac{\ln\omega_{\vec k,\sigma}}{k_z}\xi_{\vec k,\sigma}\cos\vartheta_{\vec k}.
 \end{gather}

\subsection{Quantum effects}

One can further compute the spin current in the quantized form by replacing the Rayleigh-Jeans distribution $n_{\vec k,\sigma}^0$ with the Bose-Einstein distribution $n^\text{BE}_{\vec k,\sigma}$.
For simplicity we restrict this discussion to the case of a temperature step as the findings are expected to be qualitatively similar for the thermal gradient.

In quantum statistics, the magnon occupation number is no longer linear in the temperature, hence, the base temperature will be relevant. 
The resulting spin current \eqref{eq:j_k_sigma} emitted from our temperature step should thus be proportional to 
\begin{gather}
n_{\vec k,\sigma} = n_{\vec k, \sigma}^\text{BE}[k_\text B(T_0 + \Delta T)] - n_{\vec k, \sigma}^\text{BE}[k_\text BT_0].
\end{gather}

\begin{figure}[!t]
	\centering
	\includegraphics[width = \columnwidth]{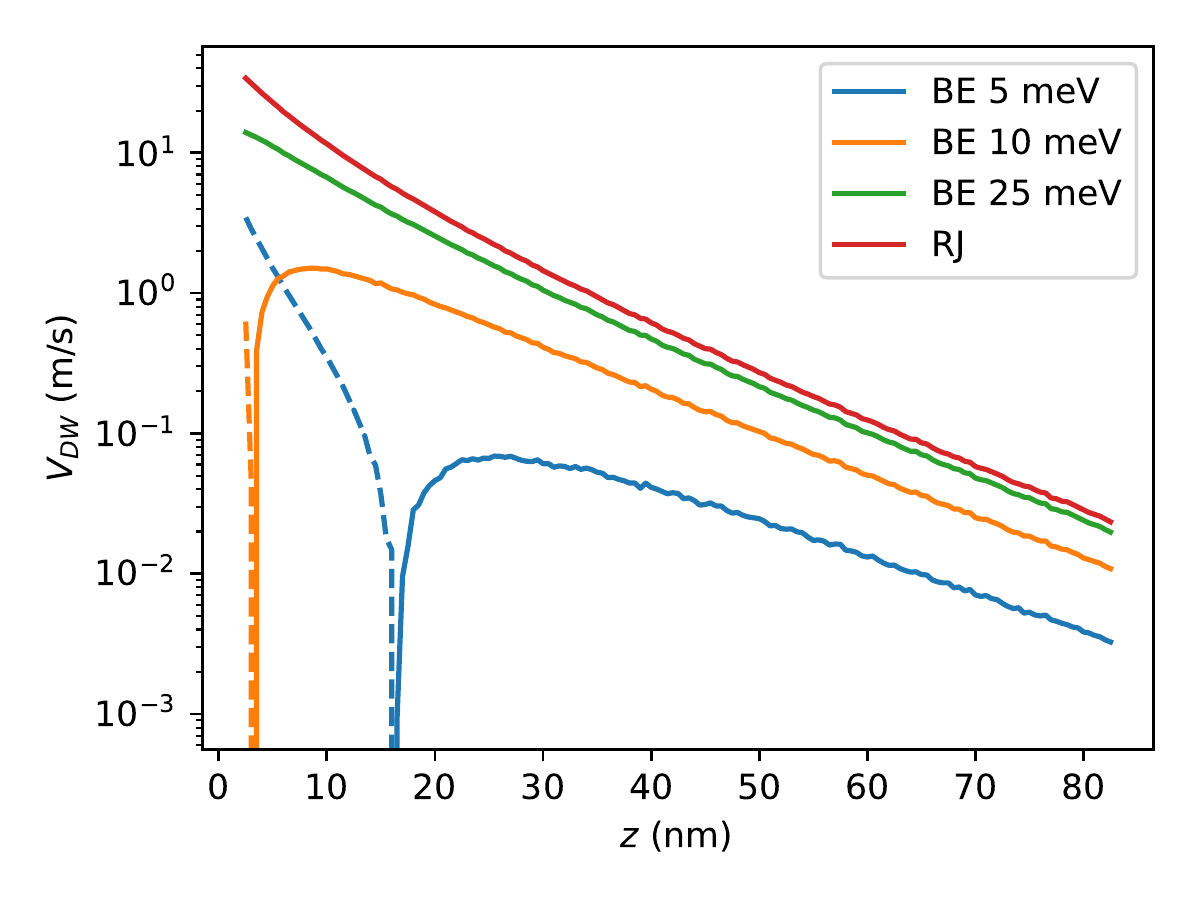}
	\caption{
	Comparison of the semiclassical spin current emitted from a $k_\text B\Delta T=\SI{1}{\milli\electronvolt}$ temperature step, derived via the Rayleigh-Jeans distribution (RJ; red line) and the quantum statistical derivation based on the Bose-Einstein distribution (BE) for different base temperatures $T_0$.
	At very low temperature the spin current is dominated by the low-frequency branch, since the high-frequency magnons are still frozen out.
	Thus the sign of the net-spin current is reversed in the vicinity of the source, indicated by the dashed line.
	}
	\label{fig:SM:QM}
\end{figure}

Figure~\ref{fig:SM:QM} shows the \ac{DW} velocity corresponding to a thermal spin current calculated with the correct quantum statistics for a set of base temperatures $k_\text BT_0$ (the Curie temperature for the given exchange constants is $T_\text C = \SI{616}{\kelvin}$). 
At very low temperature only the lowest magnon energies will be occupied, i.e. the low-frequency branch of the dispersion will dominate the spin transport, despite the low propagation length.
Moreover it is implied that real systems can exhibit a sign change of the net-spin current $J$ at very low temperature.
Thus, for $u_\text W\to0$, the \ac{DW} velocity $V_\text{DW} = Ja^3 / l_\text{fu}$ [Eqs.~\eqref{eq:adiabatic} and \eqref{eq:v-above}] not only changes sign at the compensation point where $l_\text{fu}$ changes sign, but also a second time when the temperature is sufficiently high, to populate the long-range, high-frequency magnons of the upper branch---the ones that carry negative momentum (see Fig.~\ref{fig:SM:amplitudes}).
The overall magnon current and magnon accumulation at low temperature is greatly reduced with respect to the classical case. 
However, for higher temperatures and in particular at room temperature, we qualitatively retain the semiclassical magnon current derived with the Rayleigh-Jeans distribution. 
From this we can conclude that the semiclassical treatment is sufficiently accurate for describing most experiments, which are usually carried out near room temperature with magnetic materials of similar ordering temperature~\cite{lit:RPP-Buschow}.

\section{Entropic torque in the ferrimagnet}\label{sec:SM:Schlickeiser}
The entropic or magnetothermal torque in the \ac{FI} can be defined in analogy to the \ac{FM} case \cite{lit:PRL-SRHN}.
The effective exchange stiffness for our cubical \ac{FI} is composed by the three exchange contributions $A_\text{eff}=A_\text{AA}+A_\text{AB}+A_\text{BB}$. In the molecular field approximation, for a magnetic texture along the (001) direction, these can be written as \cite{lit:PRB-Atx-etal} 
\begin{gather}
	A_\text{AA} = \frac{2J_\text{AA}}{a}m_\text A^2,~\nonumber
	A_\text{BB} = \frac{2J_\text{BB}}{a}m_\text B^2,\\
	\text{and}~~A_\text{AB} = -\frac{J_\text{AB}}{2a}m_\text Am_\text B.
\end{gather}
The different numerical factors come from the symmetry of the shells which is fcc for the ferromagnetic exchanges $A_\text{AA}$ and $A_\text{BB}$ (eight neighbors with $\Delta z=\pm1$), and simple cubic for the antiferromagnetic exchange $A_\text {AB}$ (two neighbors with $\Delta z=\pm1$), see Fig.~\ref{fig:curie}.
Their temperature dependence is hereby assumed to be well approximated by the mean field expressions $A_{ij}/A_{ij}(0)=m_im_j$.
It should be noted that although we chose exchange parameters with $J_\text{AA}\gg J_\text{BB}$ and thus $J_\text {BB}$ is not significantly affecting the magnetic ordering ($T_\text C$ for instance), it does add a non-negligible contribution to the entropic torque due to the faster demagnetization of $m_\text B$ compared to $m_\text A$.
The temperature dependence of the equilibrium magnetizations $m_i(T)$ is taken from the data in Fig.\ref{fig:curie}. 
We find $\partial m_\text A/\partial T\approx\SI{-4.87e-4}{\per\kelvin}$ for the strongly coupled sublattice A and $\partial m_\text B/\partial T\approx \SI{-2.15e-3}{\per\kelvin}$ for the weakly coupled lattice B.
The temperature derivatives which we obtained for the three exchange stiffness contributions are summarized in Tb.~\ref{tab:exchange_stiffness}.

\begin{table}[!h]
\caption{Low-temperature exchange stiffnesses and their temperature derivatives in \SI[per-mode=symbol]{e-11}{\joule\per\meter} and \SI[per-mode=symbol]{e-14}{\joule\per\meter\per\kelvin}, respectively.}
\begin{ruledtabular}
\begin{tabular}{crrrr}
	& AA & BB & AB & sum\\\hline
	$ A_{ij}$  & \num{2.05} & \num{0.064} & \num{0.192} & \num{2.31}\\
	$\partial A_{ij}/\partial T$ & \num{-2.00} & \num{-0.275} & \num{-0.507} & \num{-2.78}
\end{tabular}
\end{ruledtabular}
\label{tab:exchange_stiffness}
\end{table}

The effective magnetocrystalline anisotropy density is $K_\text{eff}^{zz} = (d_\text A^z + d_\text{B}^z)/2a^3 = \SI{5.13e6}{\joule\per\cubic\meter}$. 
This value is chosen rather high in order to (i) keep the \ac{DW} width and hence the required computation grid small and (ii) to reduce the characteristic time scale of the \ac{DW} acceleration which is proportional to the wall width $\Delta_\text{DW}$ [see Eq.~\eqref{eq:dot-Phi}].
In our simulations we observe a \ac{DW} width of about \SIrange[range-phrase=\text{~to~}]{1.6}{2.2}{\nano\meter} (depending on temperature) which is in good agreement with the theoretical prediction of  $\Delta_\text{DW} =\sqrt{A_\text{eff} / 2 K_\text{eff}^{zz}} = \SI{1.50}{\nano\meter}$

\section{Computation of steady state domain wall dynamics}\label{sec:SM:steady-state}

Due to the different time scales involved in the ferrimagnetic \ac{DW} dynamics, determining the steady-state velocity, precession, and tilting is challenging. 
On the one hand, simulation time should be as short as possible, in order to minimize the thermal drift, i.e. the error margins of the temperature, but at the same time one has to assure the simulation time is sufficiently long for the wall to reach its steady-state motion.

For the data in Fig.~\ref{fig:walkers} the steady-state dynamics were determined as follows: 
below Walker breakdown, we simulated a fixed amount of time of \SI{320}{\pico\second}, \SI{128}{\pico\second}, \SI{128}{\pico\second}, and \SI{384}{\pico\second} for the panels (a)-(d), respectively. 
These numbers reflect the acceleration time scales at the different base temperatures.
The first \SI{25}{\percent} of this simulation time was hereby discarded in order to reach the steady-state velocity (and tilting), the other \SI{75}{\percent} were used for computing the time average of $V_\text{DW}$ displayed in Fig.~\ref{fig:walkers}.
The Walker breakdown was defined by the wall angle tilting by more than \SI{45}{\degree} plus a five degree error margin, in order to account for the diverging wall precession time at exactly the Walker threshold.
Above the torque compensation point, panel (d), or very close to the Walker thresholds, the \ac{DW} precession is slow and the precession period can be several hundreds of picoseconds.
In this case, the steady-state velocity was time-averaged over only one \SI{180}{\degree}-rotation to keep thermal drift as low as possible and ensure a well-defined temperature.
For the faster precessing walls in panels (a) and (b), the precession period can be as low as few tens of picoseconds, hence, we simulated several precession periods to improve the signal to noise ratio.
In this case, the time average over \SI{256}{\pico\second} and \SI{128}{\pico\second} of simulation time was taken, respectively (rounded down to the next integer number of \SI{180}{\degree} rotations).

In Fig.~\ref{fig:temperature-sweep} (a) and (c), the steady-state velocity was determined by fitting an exponential function $\sim V_\text{DW}(1-\mathrm e^{-t/\tau})$ to the velocity data using a \SI{320}{\pico\second} simulation time.
This procedure was not applicable in Fig.~\ref{fig:walkers}, since the corresponding fits would not converge properly, especially for the lowest temperature gradients.

Finally, for the data in Fig.~\ref{fig:temperature-sweep} (b) and (d) we simply took the time average over a comparably short simulation time of \SI{128}{\pico\second}, due to the lack of inertia in the uniaxial \ac{FI}.

\bibliography{literature}

\begin{thebibliography}{83}
\bibitem [{foo({\natexlab{a}})}]{foot:grid-size}%
  \BibitemOpen
  \bibinfo {note} {For the simulations of the two-sublattice \ac{FM}, presented
  in Fig.~\ref{fig:dynamics}, larger systems of up to
  $\num{96}\times\num{96}\times{1536}$ spins were used, due to the
  significantly slower dynamics and subsequently larger distances the \ac{DW}
  propagates until a steady-state motion is reached.}\BibitemShut {Stop}%
\bibitem [{foo({\natexlab{b}})}]{foot:wall-width}%
  \BibitemOpen
  \bibinfo {note} {The signed wall width is necessary to avoid an unphysical
  discontinuity in the wall precession.}\BibitemShut {Stop}%
\end{thebibliography}

\end{document}